\documentclass[a4paper,11pt]{article}
\pdfoutput=1 

\usepackage{jheppub}
\usepackage[T1]{fontenc} 
\usepackage{graphicx}
\usepackage{soul}
\usepackage{amsmath}
\usepackage{amssymb}
\usepackage{subcaption}
\usepackage{braket}
\usepackage{xcolor}
\usepackage{lineno}
\usepackage{cancel}

\def\ba{\begin{eqnarray}}
\def\ea{\end{eqnarray}}
\def\be{\begin{equation}}
\def\ee{\end{equation}}

\def\qtraj{\texttt{QTraj} }

\title{Heavy quarkonium dynamics at next-to-leading order in the binding energy over temperature}
\preprint{TUM-EFT 169/22}

\author[a,b,c]{Nora Brambilla,}
\author[d]{Miguel \'{A}ngel Escobedo,}
\author[e]{Ajaharul Islam,}
\author[e]{Michael Strickland,}
\author[e]{Anurag Tiwari,}
\author[a]{Antonio Vairo,}
\author[a]{Peter Vander Griend}

\affiliation[a]{Physik-Department, Technische Universit\"{a}t M\"{u}nchen, James-Franck-Str.~1, 85748 Garching,
Germany}
\affiliation[b]{Institute for Advanced Study, Technische Universit\"{a}t M\"{u}nchen, Lichtenbergstrasse 2 a, 85748
Garching, Germany}
\affiliation[c]{Munich Data Science Institute, Technische Universit\"{a}t M\"{u}nchen, Walther-von-Dyck-Strasse 10, 85748 Garching, Germany}
\affiliation[d]{Instituto Galego de F\'{i}sica de Altas Enerx\'{i}as (IGFAE), Universidade de Santiago de Compostela. E-15782, Galicia, Spain}
\affiliation[e]{Department of Physics, Kent State University, Kent, OH 44242, United States}

\emailAdd{nora.brambilla@ph.tum.de}
\emailAdd{miguelangel.escobedo@usc.es}
\emailAdd{aislam2@kent.edu}
\emailAdd{mstrick6@kent.edu}
\emailAdd{anurag.tiwari128@gmail.com}
\emailAdd{vandergriend@tum.de}
\emailAdd{antonio.vairo@tum.de}

\abstract{
Using the potential non-relativistic quantum chromodynamics (pNRQCD) effective field theory, we derive a Lindblad equation for the evolution of the heavy-quarkonium reduced density matrix that is accurate to next-to-leading order (NLO) in the ratio of the binding energy of the state to the temperature of the medium.
The resulting NLO Lindblad equation can be used to more reliably describe heavy-quarkonium evolution in the quark-gluon plasma at low temperatures compared to the leading-order truncation.
For phenomenological application, we numerically solve the resulting NLO Lindblad equation using the quantum trajectories algorithm. 
To achieve this, we map the solution of the three-dimensional Lindblad equation to the solution of an ensemble of one-dimensional Schr\"odinger evolutions with Monte-Carlo sampled quantum jumps.  
Averaging over the Monte-Carlo sampled quantum jumps, we obtain the solution to the NLO Lindblad equation without truncation in the angular momentum quantum number of the states considered.  
We also consider the evolution of the system using only the complex effective Hamiltonian without stochastic jumps and find that this provides a reliable approximation for the ground state survival probability at LO and NLO. 
Finally, we make comparisons with our prior leading-order pNRQCD results and experimental data available from the ATLAS, ALICE, and CMS collaborations.
}

\begin{document}

\maketitle
\flushbottom

\section{Introduction}
\label{sect:intro}
In the pioneering work of Matsui and Satz \cite{Matsui:1986dk}, heavy quarkonium suppression was proposed as a signal of the formation of a quark gluon plasma (QGP). 
Since then, quarkonium suppression studies have been an important part of the heavy-ion program of several experimental facilities. 
For recent studies at LHC and RHIC experiments, see refs.~\cite{Acharya:2020kls,ATLAS5TeV,Sirunyan:2018nsz,CMS:2017ycw,CMS:2020efs,ALICE:2019pox,STAR:2013kwk,PHENIX:2014tbe,STAR:2016pof,CMSupsilonQM2022}.
In ref.~\cite{Matsui:1986dk}, Matsui and Satz conjectured that the screening of chromoelectric fields generated by the medium at distances proportional to the inverse of the Debye mass induces quarkonium dissociation and consequently quarkonium suppression in medium. 
In other words, the heavy quark-antiquark potential acquires a screening factor $\exp({- m_D r}) $, where $r$ is the quark-antiquark distance and $m_D$ the Debye mass which is proportional to the temperature.
Within this screening picture, quarkonia states of different radii dissociate at different temperatures providing a QGP thermometer.

The last decades have seen a change in this screening driven dissociation  paradigm.
A seminal  perturbative calculation of the heavy-quark potential by Laine, et. al., found that in the screening regime $r\sim m_D^{-1} $, in addition to Debye screening of the real part of the potential, a large imaginary contribution to the  potential also arises \cite{Laine:2006ns}. 
Perturbative calculations based on nonrelativistic effective theories of QCD at finite temperature \cite{Brambilla:2008cx,Beraudo:2007ky,Escobedo:2008sy,Brambilla:2010vq,Brambilla:2011sg,Brambilla:2013dpa} and perturbative QCD resummation \cite{Dumitru:2009fy} confirmed this result. 
Different temperature regimes have been treated using pNRQCD; outside the screening regime, different patterns of finite temperature corrections to the potential emerge, with all possibilities possessing imaginary contributions. 
 
 At temperatures relevant in current heavy ion collision experiments, the imaginary part of the potential dominates the screening contributions, and it is thus the imaginary contributions to the potential, rather than screening effects, which trigger in-medium quarkonium dissolution and suppression.
 The real part of the potential may be approximated by the quark-antiquark gauge invariant free energy in some cases \cite{Berwein:2017thy,Bazavov:2018wmo}.
 The imaginary part of the potential has its origin in two effects: Landau damping 
 (related to parton dissociation  \cite{Brambilla:2013dpa}) and singlet to octet transitions (an effect 
 particular to QCD \cite{Brambilla:2008cx} related to the gluodissociation 
 process \cite{Brambilla:2011sg}).
 Nonperturbative lattice QCD and classical real-time lattice measurements of the imaginary part of the potential were recently performed in refs.~\cite{Rothkopf:2011db,Rothkopf:2019ipj,Petreczky:2018xuh,Bala:2019cqu} and \cite{Laine:2007qy,Lehmann:2020fjt,Boguslavski:2020bxt}, respectively.
 The imaginary part of the potential indicates that quarkonium has a medium-induced decay width and, therefore, a finite lifetime. 
 There are situations in which the imaginary part of the potential is as large as the real part and cannot be treated as a perturbation, and, as mentioned, in the small coupling limit, this happens at a temperature smaller than the screening temperature. 
 Extensive phenomenological studies on the impact of the imaginary part of the static potential were performed in refs.~\cite{Strickland:2011mw,Strickland:2011aa,Krouppa:2015yoa,Islam:2020bnp,Islam:2020gdv}.

In addition to the thermal width and screening effects, an accurate description of in-medium quarkonium evolution must account for the recombination process \cite{Braun-Munzinger:2000csl,Thews:2000rj} in which an unbound heavy quark-antiquark pair recombines inside the medium to form a new bound state. 
One may distinguish between correlated recombination and uncorrelated recombination. 
The former describes the recombination of heavy quarks which initially belonged to a bound state which dissociated; the latter increases with the number of heavy quarks created in the collision and is, therefore, believed to have a significant role in charmonium production at the LHC.

It is challenging to find a theoretical description that can consistently describe all these effects. 
On the one hand, when considering screening, we aim to understand if bound state formation is possible.
This requires a quantum description. 
On the other hand, decays and recombination require consideration of transitions in which the heavy quark pairs change their color state due to the interaction with the medium. 
Therefore, we need a formalism that is quantum and describes exchanges of momentum, color, energy, etc., with the medium. 
The Open Quantum Systems (OQS) formalism \cite{Breuer:2002pc} offers a natural possibility to achieve such a description. 
Within this framework, we regard the heavy quark-antiquark state as an open quantum system interacting with an environment (the medium). 
The key object here is the reduced density matrix, obtained from the full density matrix by performing a trace over the degrees of freedom of the environment. 
The resulting equation describing the evolution of the reduced density matrix is called the master equation. 
In recent years, many works have addressed quarkonium suppression using this formalism \cite{Akamatsu:2011se,Akamatsu:2020ypb,Akamatsu:2014qsa,Miura:2019ssi,Brambilla:2016wgg,Brambilla:2017zei,Brambilla:2019tpt,Sharma:2019xum,Blaizot:2015hya,Blaizot:2017ypk,Blaizot:2018oev,Blaizot:2021xqa,Yao:2018nmy,Yao:2020xzw,Yao:2020eqy,Yao:2021lus,Katz:2015qja}. 
In this work, we focus on an approach combining the OQS framework with the use of effective field theories (EFTs) to exploit the non-relativistic nature of heavy quarkonium.

Due to its nonrelativistic nature, i.e., $v\ll 1$, where $v$ is the velocity of the heavy quarks around the center of mass, heavy quarkonium possesses a number of widely separated energy scales. 
At zero temperature, we identify: the heavy quark mass $M$, the inverse of the typical radius $1/r\sim Mv$, and the binding energy $E\sim Mv^2$.
An additional relevant scale is $\Lambda_{\rm QCD}$ at which nonperturbative effects become dominant.
Nonperturbative effects and the breaking of the weak coupling expansion when $\alpha_{\rm s}/v \sim 1$ challenge the use of perturbation theory to describe quarkonium physics, whereas at the same time lattice QCD computations struggle to simultaneously accomodate widely separated scales on sufficiently large and fine lattices.
Effective field theories address and manage these problems.

An EFT is a quantum field theory that gives the same results as the more general theory at any order of the expansion in a small parameter but is  restricted to a reduced kinematical region. 
The Lagrangian of an EFT has an infinite number of terms. 
However, given a desired precision, only a finite number of them are needed. 
Nonrelativistic QCD (NRQCD) \cite{Caswell:1985ui,Bodwin:1994jh} is an EFT that is equivalent to QCD for energies much smaller than $M$. 
It is useful to  go a step further and use potential NRQCD (pNRQCD) \cite{Pineda:1997bj,Brambilla:1999xf,Brambilla:2004jw}, an EFT equivalent to QCD for energies much smaller than $Mv$.
Such a theory makes it explicit that the appropriate zeroth order problem for the quarkonium bound state dynamics amounts to solving a Schr\"odinger equation and gives a field theoretical definition of the potentials as Wilson coefficients.
In pNRQCD, the degrees of freedom are not heavy quarks and antiquarks but the color singlet and octet states that one can form with pairs of them. 
At leading order (LO) in the power counting, the evolution of the singlet field is given by a Schrödinger equation. 
However, at higher orders, interactions with gauge fields with energy smaller than $Mv$ become important and contain retardation effects.
Systematic calculations of energy levels, decays, and transitions are made  possible by the EFT and the power counting allows one to estimate the uncertainty in the prediction of any observable.

Non-relativistic EFTs can also be used to study quarkonium at finite temperature \cite{Brambilla:2008cx,Escobedo:2008sy,Brambilla:2010vq,Brambilla:2011sg,Brambilla:2013dpa}. 
In this case, we must consider also the energy scales induced by the medium, e.g., the temperature $T$. 
In particular, pNRQCD can be realized in medium and combined with the OQS framework to derive a master equation governing the evolution of in-medium heavy quarkonium. 
References~\cite{Brambilla:2016wgg,Brambilla:2017zei,Brambilla:2019tpt} analyzed the regime $Mv \gg T$.
In this case, the matching between NRQCD and pNRQCD is unaffected by the medium, and one can perform the calculation of the relevant diagrams contributing to the in-medium evolution 
using the~\mbox{$T=0$} pNRQCD Lagrangian.
Using the closed path real-time formalism, the authors of refs.~\cite{Brambilla:2016wgg,Brambilla:2017zei} derived a master equation for the in-medium heavy quarkonium color singlet and color octet density matrices. 
We emphasize that the theory, methods, and results are fully quantum, nonabelian, and heavy quark number conserving. 
With the further assumption $T\gg E$ the master equation could be cast in Lindblad form \cite{Gorini:1975nb,Lindblad:1975ef} at leading order in the $E/T$ expansion. 
In a series of works \cite{Brambilla:2016wgg,Brambilla:2017zei,Brambilla:2020qwo,Brambilla:2021wkt}, a number of phenomenological predictions were made for bottomonium suppression in the temperature regime $Mv \gg T \gg E$; physically, this amounts to using bottomonium states of small radius as probes of a strongly coupled QGP.
In the most recent works \cite{Brambilla:2020qwo,Brambilla:2021wkt}, the \texttt{QTraj} code of ref.~\cite{Omar:2021kra}, which implements the Monte Carlo quantum trajectories algorithm \cite{Dalibard:1992zz}, was used to solve the Lindblad equation at substantially reduced computational cost compared to previous works, while simultaneously implementing a realistic medium evolution by coupling to a state of the art 3D viscous hydrodynamics code.
In this way, one obtains a good description of existing experimental results on bottomonium suppression and in-medium observables including the nuclear modification factor $R_{AA}$ and the elliptic flow $v_2$. 
We note that these results do not require, in principle,  the fixing of free parameters to experimental data: the strong coupling $\alpha_{\rm s}$ we take from the Particle Data Group~\cite{Zyla:2020zbs}, for the bottom mass, we work in the $1S$ scheme where $m_{b}=m_{\Upsilon(1S)}/2$ with the $\Upsilon(1S)$ mass also from the Particle Data Group, while the nonperturbative transport coefficients $\kappa$ and $\gamma$ describing the strongly coupled QGP may be taken from direct or indirect lattice measurements, as they are expressed in terms of gauge invariant correlators of chromoelectric fields.

In the previous phenomenological studies \cite{Brambilla:2020qwo,Brambilla:2021wkt}, the requirement that the inequality $T\gg E$ be fulfilled over the course of the evolution necessitated the termination of the coupling to the medium at $T_{f}=250$ MeV. 
In the present paper, we aim to improve on this by considering the next-to-leading order (NLO) corrections in the $E/T$ expansion.
Previous computations in other models \cite{Blaizot:2018oev,Miura:2019ssi,DeBoni:2017ocl} indicate that some of the NLO corrections are related to the drag force on a particle moving in the medium. 
In this way, heavy quarkonium loses energy to the medium thus bringing it closer to a thermal distribution. 
We note that in the Abelian case considered in refs.~\cite{Blaizot:2018oev,Miura:2019ssi,DeBoni:2017ocl} the NLO corrections in the $E/T$ expansion bring the system to thermal equilibrium.

In this paper, we compute the first corrections in the $E/T$ expansion to the master equation of pNRQCD in the region $Mv \gg T \sim gT \gg E$ and explore their phenomenological impact. 
As explained, we work in QCD, in a fully quantum and nonabelian setup, considering the case of quarkonium evolution in a strongly coupled medium. 
Moreover, we implement a realistic hydrodynamic evolution.  
Phenomenological results at NLO  in such a case have never been obtained before. 
As such, our results may be relevant not only to the study of the nonequilibrium evolution of quarkonium in medium but also to the study of the approach to equilibrium.  
We plan to address the second point in a future publication and here focus on the first point, the phenomenological impact of NLO corrections.

This paper is organized as follows. 
In sec.~\ref{sec:nlolindblad}, we extend our open quantum system (OQS) treatment of in-medium heavy quarkonium to include corrections at NLO in the $E/T$ expansion and construct an NLO evolution equation of Lindblad form for the heavy quarkonium reduced density matrix.   
Taking advantage of the isotropy of the problem, we expand the Lindblad equation in spherical harmonics to rewrite the 3-dimensional evolution in terms of a 1-dimensional evolution; we discuss the form of the collapse operators and in-medium widths in this expansion.
In sec.~\ref{sec:qtrajnlo}, we discuss the implementation of the quantum trajectories method at this order. 
In sec.~\ref{sec:results}, we present  our results.
On the theoretical side, we discuss the effect of the NLO terms on the in-medium widths and survival probabilities and the effect of quantum jumps.
On the phenomenological side, we present our results for the nuclear modification factor $R_{AA}$ of the $\Upsilon(1S)$, $\Upsilon(2S)$, and $\Upsilon(3S)$ as a function of number of participating nucleons $N_{\text{part}}$ and transverse momentum $p_{T}$. 
We extend the evolution of the Linblad equation down to $T_{f}=190$ MeV and still obtain a satisfactory description of the experimental data. 
We comment on the dependence and features of our results in relation to the present uncertainty of the lattice determinations of the transport coefficients $\kappa$ and $\gamma$.
We conclude and provide an outlook on future prospects in sec.~\ref{sec:conclusions}.

\section{Derivation of the NLO Lindblad equation}
\label{sec:nlolindblad}
In this section, we expand the evolution equation obeyed by the heavy quarkonium reduced density matrix to include terms of order $E/T$ and derive a Lindblad equation accurate at this order.
In this way, we go beyond the analysis at leading order in $E/T$ presented in refs.~\cite{Brambilla:2016wgg,Brambilla:2017zei,Brambilla:2020qwo,Brambilla:2021wkt}.

\subsection{The pNRQCD master equation}\label{sec:master_equation}

The evolution of the reduced density matrix of heavy quarkonium in the regime $Mv \gg T$ was derived in refs.~\cite{Brambilla:2016wgg,Brambilla:2017zei}.
If one does not make any assumption on the relation between $T$ and $E$, then the evolution equations that one obtains are 
\begin{align}
	\frac{d\rho_{s}(t)}{dt} &= -i[h_{s}, \rho_{s}(t)] - \Sigma_{s} \rho_{s}(t) - \rho_{s}(t) \Sigma^{\dagger}_{s} + \Xi_{so}(\rho_{o}(t)),\\
	\frac{d\rho_{o}(t)}{dt} &= -i[h_{o}, \rho_{o}(t)] - \Sigma_{o} \rho_{o}(t) - \rho_{o}(t) \Sigma_{o}^{\dagger}
	+ \Xi_{os}(\rho_{s}(t)) + \Xi_{oo}(\rho_{o}(t)),
\end{align}
where, assuming the in medium correlators to be isotropic and time-translation invariant
and observing the evolution for a time $t$ larger than any other length scale in the problem,\footnote{
These assumptions account for the difference between eq.~\eqref{eq:auv_operator} and eq.~(D.2) of ref.~\cite{Brambilla:2017zei}.}
\begin{align}
	\Sigma_{s} &= r_{i} A_{i}^{so \dagger}, \\
	\Sigma_{o} &= \frac{1}{N_{c}^{2}-1} r_{i} A_{i}^{os \dagger}
	+ \frac{N_{c}^{2}-4}{2(N_{c}^{2}-1)}r_{i}A_{i}^{oo \dagger}, \\
	\Xi_{so}(\rho_{o}(t)) &= \frac{1}{N_{c}^{2}-1} \left( A_{i}^{os \dagger} \rho_{o}(t) r_{i} + r_{i} \rho_{o}(t) A_{i}^{os}\right), \\
	\Xi_{os}(\rho_{s}(t)) &= A_{i}^{so \dagger} \rho_{s}(t) r_{i} 
	+ r_{i} \rho_{s}(t) A_{i}^{so}, \\
	\Xi_{oo}(\rho_{o}(t)) &= \frac{N_{c}^{2}-4}{2(N_{c}^{2}-1)}\left(
	A_{i}^{oo \dagger} \rho_{o}(t) r_{i} + r_{i} \rho_{o}(t) A_{i}^{oo}\right),
\end{align}
with
\begin{align}
	A_{i}^{uv} &= \frac{g^{2}}{6N_{c}} \int^{\infty}_{0} \text{d}s\, e^{-i h_{u}s} r_{i} e^{i h_{v} s} 
	\langle \tilde{E}^{a}_j(0, \vec{0}) \tilde{E}^{a}_j(s, \vec{0})\rangle,\label{eq:auv_operator}\\
	\tilde{E}^{a}_i(s, \vec{0}) &= \Omega(s)^\dagger E^{a}_i(s, \vec{0}) \Omega(s),\\
	\Omega(s) &= \text{exp}\left[  -ig \int_{-\infty}^{s} \text{d}s' A_{0}(s', \vec{0}) \right].
\end{align}
$h_{u,v}$ is the singlet or octet Hamiltonian: 
$h_{s,o}=\vec{p}^{2}/M+V_{s,o}$ where $V_{s}= -C_{f}\alpha_{\rm s}(1/a_{0})/r$ and $V_{o}=\alpha_{\rm s}(1/a_{0})/2N_{c}r$, 
with $C_{f}=(N_{c}^{2}-1)/2N_{c} = 4/3$ the Casimir of the fundamental representation, $N_{c}=3$ the number of colors, and $\alpha_{\rm s}(1/a_{0})$ the strong coupling evaluated at the inverse of the Bohr radius $a_{0}$.
Details about the derivation of these results can be found in refs.~\cite{Brambilla:2017zei,Brambilla:2019tpt}. 

These equations can be written as the master equation
\begin{equation}\label{eq:master_equation}
	\frac{d\rho(t)}{dt} = -i \left[ H, \rho(t) \right] + \sum_{nm} h_{nm} \left( L_{i}^{n} \rho(t) L^{m\dagger}_{i} - \frac{1}{2} \left\{ L^{m\dagger}_{i} L_{i}^{n}, \rho(t) \right\} \right),
\end{equation}
where
\begin{equation}\label{eq:master_equation_rho_and_h}
	\rho(t) = \begin{pmatrix} \rho_{s}(t) & 0 \\ 0 & \rho_{o}(t) \end{pmatrix} ,\text{\quad} H = \begin{pmatrix} h_{s} + \text{Im}(\Sigma_{s}) & 0 \\ 0 & h_{o} + \text{Im}(\Sigma_{o}) \end{pmatrix} ,
\end{equation}
\begin{equation}\label{eq:master_equation_l0_l1}
	L_{i}^{0} = \begin{pmatrix} 0 & 0 \\ 0 & 1 \end{pmatrix}r_{i} \text{, \quad} L_{i}^{1} = \begin{pmatrix} 0 & 0 \\ 0 & \frac{N_{c}^{2}-4}{2(N_{c}^{2}-1)} A_{i}^{oo \dagger} \end{pmatrix},
\end{equation}
\begin{equation}\label{eq:master_equation_l2_l3}
	L_{i}^{2} = \begin{pmatrix} 0 & 1 \\ 1 & 0 \end{pmatrix}r_{i} \text{, \quad} L_{i}^{3} = \begin{pmatrix} 0 & \frac{1}{N_{c}^{2}-1} A_{i}^{os \dagger} \\ A_{i}^{so \dagger} & 0 \end{pmatrix},
\end{equation}
and
\begin{equation}\label{eq:metric_tensor}
	h = \begin{pmatrix} 0 & 1 & 0 & 0 \\ 1 & 0 & 0 & 0 \\ 0 & 0 & 0 & 1 \\ 0 & 0 & 1 & 0 \end{pmatrix}.
\end{equation}
If the matrix $h$ were a completely positive matrix, its eigenvalues would be positive and we could make a change of variables to bring eq.~(\ref{eq:master_equation}) into Lindblad form.
However, as the matrix $h$ is not completely positive definite, 
this manipulation is not possible. 
Any Markovian, trace preserving evolution that preserves complete positivity of the density matrix can be written as a Lindblad equation.
As eq.~(\ref{eq:master_equation}) is Markovian and trace preserving but does not preserve complete positivity of the density matrix, it follows that eq.~(\ref{eq:master_equation}) allows for negative probabilities, a feature similar to the widely used Caldeira--Leggett master equation \cite{Caldeira:1982iu}. 
We note that complete positivity is necessary for use of the quantum trajectories method.

\subsection{Lindblad equation to leading-order in \texorpdfstring{$E/T$}{E/T}}
\label{sec:lindbladlo}

Although, as explained above, the evolution specified in eq.~(\ref{eq:master_equation}) does not, in general, take the form of a Lindblad equation, in specific temperature regimes the operators characterizing the medium interaction simplify, and the resulting evolution equation can be rewrtitten in Lindblad form.
We consider the case $T\gg E$.
In eq.~(\ref{eq:auv_operator}), the thermal chromoelectric correlator decays rapidly at times $s> 1/T$, while the quarkonium Hamiltonian $h_{s,o}$ scales as the binding energy $E$.
The arguments of the exponentials in $A_{i}^{uv}$ thus scale like $E/T$.
In the limit $T\gg E$, the exponentials can be set to $1$, and $A_{i}^{uv}$ can be written in terms of the transport coefficients $\kappa$ and $\gamma$ 
\begin{equation}
    A_{i}^{uv} = \frac{r_{i}}{2} \left( \kappa - i \gamma \right) + \cdots,
\end{equation}
where the ellipsis indicates terms of order $E/T$ and higher; 
$\kappa$ is the heavy quarkonium momentum diffusion coefficient, and $\gamma$ is its dispersive counterpart
\begin{align}
	\kappa &= \frac{g^{2}}{6N_{c}} \int_{0}^{\infty}\text{d}s\, \left\langle \left\{\tilde{E}^{a}_i(s,\vec{0}), \tilde{E}^{a}_i(0,\vec{0})\right\}\right\rangle ,\label{eq:kappa}\\ 
	\gamma &= -i\frac{g^{2}}{6N_{c}} \int_{0}^{\infty}\text{d}s\, \left\langle \left[\tilde{E}^{a}_i(s,\vec{0}), \tilde{E}^{a}_i (0,\vec{0})\right]\right\rangle.\label{eq:gamma}
\end{align}
At leading order in $E/T$, the $L^{n}$ are not linearly independent; as a consequence, there exists a rotation of the vector $(L^{0},L^{1},L^{2},L^{3})$ that makes it orthogonal to the eigenspaces of the negative eigenvalues of the matrix $h_{nm}$ (see appendix~\ref{app:nlo_lindblad} for details).
This allows to write the master equation in Lindblad form
\begin{equation}\label{eq:lindblad_lo}
\frac{d\rho(t)}{dt} = -i \left[ H, \rho(t) \right] + \sum_{n}\left( C_{i}^{n} \rho(t) C^{n\dagger}_{i} - \frac{1}{2} \left\{ C^{n \dagger}_{i} C_{i}^{n}, \rho(t) \right\} \right),
\end{equation}
with Hamiltonian
\begin{equation}
	H = \begin{pmatrix} h_{s} + \text{Im}(\Sigma_{s}) & 0 \\ 0 & h_{o} + \text{Im}(\Sigma_{o}) \end{pmatrix},
\end{equation}
where
\begin{equation}\label{eq:mass_shift_lo}
	\text{Im}\left( \Sigma_{s} \right) = \frac{r^{2}}{2} \gamma,\quad
	\text{Im}\left( \Sigma_{o} \right) = \frac{N_{c}^{2}-2}{2(N_{c}^{2}-1)} \frac{r^{2}}{2} \gamma,
\end{equation}
and collapse operators
\begin{align}
	C_{i}^{0} =& \sqrt{\frac{\kappa}{N_{c}^{2}-1}} r_{i} \begin{pmatrix} 0 & 1 \\ \sqrt{N_{c}^{2}-1} & 0 \end{pmatrix},\label{eq:c0_lo}\\
	C_{i}^{1} =& \sqrt{\frac{\kappa(N_{c}^{2}-4)}{2(N_{c}^{2}-1)}} r_{i} \begin{pmatrix} 0 & 0 \\ 0 & 1 \end{pmatrix}  \label{eq:c1_lo}.
\end{align}
This equation was first derived and solved in refs.~\cite{Brambilla:2016wgg,Brambilla:2017zei} and further studied in refs.~\cite{Brambilla:2020qwo,Brambilla:2021wkt} using the \texttt{QTraj} code of ref.~\cite{Omar:2021kra}.

\subsection{Lindblad equation to next-to-leading order in \texorpdfstring{$E/T$}{E/T}}

In the previous subsections, we presented the master equation governing the in-medium evolution of Coulombic quarkonium and showed that in the $T\gg E$ limit it takes the form of a Lindblad equation. 
In this subsection, we consider contributions to $A_{i}^{uv}$ linear in  $E/T$ and investigate the conditions under which a Lindblad equation can still be derived. 
Making use of the relation
\begin{equation}\label{eq:nlo_kappa}
	i\frac{g^{2}}{6 N_{c}}\int_{0}^{\infty}\text{d}t \, t \, \Big \langle \tilde{E}^{a}_i(t,\vec{0})\tilde{E}^{a}_i(0,\vec{0}) \Big \rangle = \frac{\kappa}{4T},
\end{equation}
which follows from the fluctuation-dissipation theorem (see appendix~\ref{app:correlatoridentities} for a detailed derivation), we obtain corrections to eqs.~(\ref{eq:mass_shift_lo})-(\ref{eq:c1_lo}) of order $E/T$ by expanding the exponentials of eq.~(\ref{eq:auv_operator}) and retaining terms up to order $h_{u,v}$.
This results in $A_{i}^{uv}$ taking the form
\begin{align}\label{eq:nlo_medium_interaction}
	A_{i}^{uv} =& \frac{r_{i}}{2} (\kappa - i \gamma) + \kappa \left( -\frac{i p_{i}}{2MT} + \frac{\Delta V_{uv}}{4T} r_{i} \right) + \cdots,
\end{align}
where $\Delta V_{uv}=V_{u}-V_{v}$ is the difference of the $u$ and $v$ (singlet or octet) potentials and the ellipsis indicates terms of order $(E/T)^{2}$ and higher.
With these contributions to $A_{i}^{uv}$, the operators $L_i^1$ and $L_i^3$ in the master equation take the form
\begin{align}
	&L_{i}^{1} = \frac{N_{c}^{2}-4}{2(N_{c}^{2}-1)} \begin{pmatrix} 0 & 0 \\ 0 &  1 \end{pmatrix} \left[ \frac{r_{i}}{2} (\kappa + i \gamma) + \kappa \frac{i p_{i}}{2MT} \right] ,
	\label{eq:Li1NLO}\\
	&\begin{aligned} L_{i}^{3}=
	&\begin{pmatrix} 0 & \frac{1}{N_{c}^{2}-1} \\ 0 & 0 \end{pmatrix} \left[\frac{r_{i}}{2} (\kappa + i \gamma) + \kappa \left( \frac{i p_{i}}{2MT} + \frac{\Delta V_{os}}{4T} r_{i} \right) \right] \\
	&+ \begin{pmatrix} 0 & 0 \\ 1 & 0 \end{pmatrix} \left[\frac{r_{i}}{2} (\kappa + i \gamma) + \kappa \left( \frac{i p_{i}}{2MT} + \frac{\Delta V_{so}}{4T} r_{i} \right) \right],
	\end{aligned}
\end{align}
while $L_i^0$ and $L_i^2$ keep the form given in eqs.~\eqref{eq:master_equation_l0_l1} and~\eqref{eq:master_equation_l2_l3}, respectively.

In contrast to the case at order $(E/T)^{0}$ considered in subsection~\ref{sec:lindbladlo}, the above master equation cannot be written in Lindblad form as the $L^{n}_i$ are linearly independent.
However, the vector $(L^{0}, L^{1}, L^{2}, L^{3})$ can be rotated in such a way that only components of order $E/T$ project on the eigenspaces of the negative eigenvalues of the matrix $h_{nm}$.
Discarding these components amounts to neglecting terms in the evolution equation of order $(E/T)^{2}$.
Hence, we can still write a Lindblad equation that is accurate at order $E/T$
\begin{equation}
\frac{d\rho(t)}{dt} = -i \left[ H, \rho(t) \right] + \sum_{n=0}^1
\left( C_{i}^{n} \rho(t) C^{n\dagger}_{i} - \frac{1}{2} \left\{ C^{n \dagger}_{i} C_{i}^{n}, \rho(t) \right\} \right),
\label{eq:Lindblad}
\end{equation}
with Hamiltonian
\begin{equation}
	H = \begin{pmatrix} h_{s} + \text{Im}(\Sigma_{s}) & 0 \\ 0 & h_{o} + \text{Im}(\Sigma_{o}) \end{pmatrix},
\end{equation}
where
\begin{equation}\label{eq:self_energies_im}
	\text{Im}\left( \Sigma_{s} \right) = \frac{r^{2}}{2} \gamma +\frac{\kappa}{4MT} \{r_{i}, p_{i}\} \text{, \quad} 
	\text{Im}\left( \Sigma_{o} \right) = \frac{N_{c}^{2}-2}{2(N_{c}^{2}-1)} \left( \frac{r^{2}}{2} \gamma +\frac{\kappa}{4MT} \{r_{i}, p_{i}\} \right),
\end{equation}
and collapse operators
\begin{align}
	&\begin{aligned}\label{eq:c0}
	    C_{i}^{0} =& \sqrt{\frac{\kappa}{N_{c}^{2}-1}} \begin{pmatrix} 0 & 1 \\ 0 & 0 \end{pmatrix} \left(r_{i} + \frac{i p_{i}}{2MT} +\frac{\Delta V_{os}}{4T}r_{i} \right) \\ 
	&+ \sqrt{\kappa} \begin{pmatrix} 0 & 0 \\ 1 & 0 \end{pmatrix} \left(r_{i} + \frac{i p_{i}}{2MT} +\frac{\Delta V_{so}}{4T}r_{i} \right),
	\end{aligned}\\
	&C_{i}^{1} = \sqrt{\frac{\kappa(N_{c}^{2}-4)}{2(N_{c}^{2}-1)}} \begin{pmatrix} 0 & 0 \\ 0 & 1 \end{pmatrix} \left(r_{i} + \frac{i p_{i}}{2MT} \right).\label{eq:c1}
\end{align}
For details, see appendix~\ref{app:nlo_lindblad}.

At the price of introducing some spurious terms at order $(E/T)^{2}$, we have derived a Lindblad equation accurate up to order $E/T$ and can thus make use of the quantum trajectories algorithm to solve it. Note that similar strategies have been used to improve the Caldeira--Leggett model \cite{diosi1993high,diosi1993calderia,gao1997dissipative,vacchini2000completely}. 
We can understand the anti-commutator term $\{r_{i}, p_{i}\}$ multiplying $\kappa$ in $\text{Im}\left( \Sigma_{s,o} \right)$ as a drag force. 
This identification is clear if one uses the procedure described in ref.~\cite{Blaizot:2017ypk} to obtain a Langevin equation from eq. (\ref{eq:Lindblad}).

\subsection{Lindblad equation to NLO in \texorpdfstring{$E/T$}{E/T} in the spherical basis}
The six collapse operators in Eqs.~(\ref{eq:c0}) and (\ref{eq:c1}) (along with the Hamiltonian $H$) encode the full 3-dimensional evolution of Coulombic quarkonium of binding energy $E$ propagating in a thermal medium of temperature $T$ in the regime $T\gg E$.\footnote{
In practice, adding $E/T$ corrections serve to relax the parametric  
condition $T\gg E$ to a milder $T \gtrsim E$ in the phenomenological applications of 
sec.~\ref{sec:results}.
}
The collapse operators implement transitions between quarkonium states of different color and angular momentum.
This can be made manifest in the angular momentum sector by projecting the density matrix onto the spherical harmonics $Y_{lm}$
\begin{equation}
    \rho^{lm;l'm'} = \int d\Omega\, d\Omega'\, Y_{lm}\, \rho\, Y_{l'm'}.
\end{equation}
As the plasma is isotropic, i.e., there is no preferred direction in space, only diagonal elements $l=l'$ and $m=m'$ are nonzero; furthermore, all information can be encoded in
\begin{equation}
    \rho^{l} = \sum_{m} \rho^{lm;lm}.
\end{equation}
We thus project the Lindblad equation onto the spherical harmonics and sum over the magnetic quantum number $m$.
After this projection, the Lindblad equation is an infinite dimensional matrix equation, which is the tensor product of the $2\times 2$ color space, the infinite dimensional angular momentum space, and the radial wave function.
The density matrix is diagonal and takes the form
\begin{equation}
    \rho = \begin{pmatrix} 
        \rho^{0}_{s} & 0 & \hdots & 0 & 0 & \cdots \\
        0 & \rho^{1}_{s} & \hdots & 0 & 0 & \hdots \\
        \vdots & \vdots & \ddots & \vdots & \vdots \\
        0 & 0 & \hdots & \rho^{0}_{o} & 0 & \hdots \\
        0 & 0 & \hdots & 0 & \rho_{o}^{1} & \hdots \\
        \vdots & \vdots & & \vdots & \vdots & \ddots
    \end{pmatrix},
\end{equation}
with the Hamiltonian being an analogous diagonal matrix, the elements of which are given by
\begin{equation}
    h^{l}_{s,o} = -\frac{1}{M}\left(\frac{\partial^{2}}{\partial r^{2}} + \frac{2}{r} \frac{\partial}{\partial r} \right) + V_{s,o} + \frac{l(l+1)}{M r^{2}}.
\end{equation}
The collapse operators take the form
\begin{align}
    &\begin{aligned}
        C^{0} =& \sqrt{\frac{\kappa}{N_{c}^{2}-1}} \begin{pmatrix} 0 & 1 \\ 0 & 0 \end{pmatrix} \otimes 
        \left( O^-
        \sqrt{\frac{l}{2l+1}} C^{\downarrow}_{o\to s} 
        + O^+
        \sqrt{\frac{l+1}{2l+1}} C^{\uparrow}_{o\to s}\right) \\
        &+ \sqrt{\kappa} \begin{pmatrix} 0 & 0 \\ 1 & 0 \end{pmatrix} \otimes 
        \left( O^-
        \sqrt{\frac{l}{2l+1}} C^{\downarrow}_{s\to o} 
        + O^+
        \sqrt{\frac{l+1}{2l+1}} C^{\uparrow}_{s\to o}\right), 
    \end{aligned}\\
    &C^{1} = \sqrt{\frac{\kappa(N_{c}^{2}-4)}{2(N_{c}^{2}-1)}} \begin{pmatrix} 0 & 0 \\ 0 & 1 \end{pmatrix} \otimes 
    \left( 
    O^-
    \sqrt{\frac{l}{2l+1}} C^{\downarrow}_{o\to o} 
    + O^+
    \sqrt{\frac{l+1}{2l+1}} C^{\uparrow}_{o\to o}\right),
\end{align}
where the $2\times 2$ matrices are responsible for color transitions and the infinite dimensional matrices $O^{\pm}_{l',l} = \delta_{l',l\pm1} $ for angular momentum transitions.
The operators $C_{u\to v}^{\uparrow, \downarrow}$ act on $\rho^{l}_{u}$ contributing to its partial width to $\rho^{l\pm 1}_{v}$; they are given explicitly by
\begin{align}
    C^{\uparrow}_{u\to v} =& r\left(1+ \frac{\Delta V_{uv}}{4T} \right) 
    + \frac{1}{2MT}\left(\frac{\partial}{\partial r} - \frac{l}{r} \right), \\
    C^{\downarrow}_{u\to v} =& r\left(1+ \frac{\Delta V_{uv}}{4T} \right) 
    + \frac{1}{2MT}\left(\frac{\partial}{\partial r} + \frac{l+1}{r} \right).
\end{align}
In this way, the 3-dimensional problem has been reduced to a 1-dimensional problem and the 6 collapse operators to 2.
For details, see appendix~\ref{app:spherical}.

This same procedure was carried out in ref.~\cite{Brambilla:2017zei} truncating the angular momentum at $l=1$. 
Since then, works based on the \texttt{QTraj} code~\cite{Omar:2021kra} have, however, made it possible to account for states of arbitrarily high angular momentum \cite{Brambilla:2020qwo,Brambilla:2021wkt}.

The decay width of the state $\rho^{l}_{u}(r)$ is given by the trace of the anti-commutator term of the Lindblad equation
\begin{equation}
    \Gamma \left[
    \rho_{u}^{l}(r)
    \right]\equiv \sum_{n=0}^1
    \text{Tr} \left[ C^{n\,\dagger} C^{n} \rho^{l}_{u}(r) \right],
\end{equation}
where on the left $\Gamma[ \rho_{u}^{l}(r) ]$ represents a functional returning the width and on the right the trace is over color, angular momentum, and position.
The partial widths with respect to each of the color channels are given by
\begin{align}
    \Gamma_{u\to v} \rho^{l}(r) =& \sum_{n=0}^1
    \text{Tr}_{c} \left[ C^{n\,\dagger} |v \rangle \langle v | C^{n} \rho^{l}_{u}(r) \right],
\end{align}
where the trace is only over color, and $|v\rangle \langle v|$ is a state of color $v$.
Computing the operators explicitly, we find
\begin{align}
    \Gamma_{s\to o} &= 
    \kappa \left\{ 
    r^{2}\left(1 + \frac{\Delta V_{so}}{4T} \right)^{2}
    - \frac{3}{2MT}
    - \frac{\Delta V_{so}}{4MT^{2}} 
    + \frac{\mathcal{D}^{2}}{4M^{2}T^{2}} \right\},\label{eq:Gamma_s_to_o}\\
    \Gamma_{o\to s} &= 
    \frac{\kappa}{N_{c}^{2}-1} \left\{ 
    r^{2} \left(1 + \frac{\Delta V_{os}}{4T} \right)^{2}
    - \frac{3}{2MT} 
    - \frac{\Delta V_{os}}{4MT^{2}} 
    + \frac{\mathcal{D}^{2}}{4M^{2}T^{2}} \right\},\\
    \Gamma_{o\to o}	&= 
    \frac{\kappa (N_{c}^{2}-4)}{2(N_{c}^{2}-1)} \left\{ 
    r^{2} 
    - \frac{3}{2MT} 
    + \frac{\mathcal{D}^{2}}{4M^{2}T^{2}} \right\},
\end{align}
where
\begin{equation}
    \mathcal{D}^2 = -\left(\frac{\partial^2}{\partial r^2} + \frac{2}{r}\frac{\partial}{\partial r} \right) + \frac{l (l+1)}{r^2} \, .
\end{equation}

\section{The quantum trajectories algorithm at NLO}
\label{sec:qtrajnlo}

The Lindblad equation given in eq.~(\ref{eq:Lindblad}) with the collapse operators of eqs.~\eqref{eq:c0} and \eqref{eq:c1} describes the evolution of the heavy quarkonium reduced density matrix at next-to-leading-order accuracy in the $E/T$ expansion.  
To solve it numerically, we make use of the quantum trajectories algorithm used to solve the LO evolution equations in refs.~\cite{Brambilla:2021wkt,Brambilla:2020qwo,Omar:2021kra}.  
In the quantum trajectories approach, one evolves a large set of independently sampled quantum evolutions of the wave function. 
Observables, e.g., the overlap with vacuum eigenstates, are computed along each sampled quantum trajectory and averaged to obtain the final predictions for the observable in question.  

Here we make use of the ``waiting time approach'' in which one evolves the wave function using the complex effective Hamiltonian until the norm squared of the wave function falls below a random number $p_{1}$ uniformly sampled in $[0,1]$, at which point additional random numbers are generated to determine the outgoing quantum numbers and corresponding collapse operator to apply to the wave function~\cite{Omar:2021kra}.  
After application of the selected collapse operator, the wave function is normalized, and the process is repeated until the simulation end time is reached.  
This method has the benefit that between quantum jumps the evolution proceeds with a fixed color state labeled by $c$ and fixed angular momentum labeled by $l$.  
For the description below, we note that as the system is isotropic we can write the Hamiltonian and collapse  operators more compactly by using the reduced wave function $u(t,r) \equiv r R(t,r)$, where the three-dimensional wave function for fixed $l$ and $m$ is given by $\psi(t,\vec{r}) = R(t,r) Y_{lm}(\theta,\phi)$.  
In subsections \ref{subsec:ci_u}, \ref{subsec:gamma_u}, and \ref{subsec:h_u}, we give the forms of the collapse operators, widths, and in-medium Hamiltonian acting on the reduced wave function; we refer to quantities acting on the reduced wave function as being in the \textit{reduced spherical representation} and denote them with an overbar.
More details on the reduced spherical representation are given in appendix~\ref{app:spherical}.
In subsection \ref{subsec:qtraj_algorithm}, we give the quantum trajectories algorithm as implemented to obtain the phenomenological results in sec.~\ref{sec:results}.
In subsection \ref{subsec:evolution_between_jumps}, we describe the procedure used to evolve the wave function between quantum jumps.

\subsection{Jump operators in the reduced spherical representation}\label{subsec:ci_u}
There are six jump operators present in the Lindblad equation. 
They represent the six possible physical transitions for the quarkonium state inside the QGP. 
A color singlet state can transition into an octet and vice versa; in addition, the octet state can also transition to another octet state. 
For each color transition, we have the two additional possibilities of an angular momentum jump up $(l \rightarrow l+1$) or down ($l \rightarrow l-1$), thus totalling six.  
When expressed as operators acting on the reduced wave function, the six jump operators are
\begin{eqnarray}
\overline{C}^{\uparrow}_{s\to o} &=& r  - \frac{N_c \alpha_{\rm s}}{8T} + \frac{1}{2 M T} \left( \frac{\partial}{\partial r}  - \frac{l+1}{r} \right) , \label{eq:cupso}\\
\overline{C}^{\downarrow}_{s\to o} &=& r  - \frac{N_c \alpha_{\rm s}}{8T} + \frac{1}{2 M T} \left( \frac{\partial}{\partial r}  + \frac{l}{r} \right) , \\
\overline{C}^{\uparrow}_{o\to s} &=& r  + \frac{N_c \alpha_{\rm s}}{8T} + \frac{1}{2 M T} \left( \frac{\partial}{\partial r}  - \frac{l+1}{r} \right) , \\
\overline{C}^{\downarrow}_{o\to s} &=& r  + \frac{N_c \alpha_{\rm s}}{8T} + \frac{1}{2 M T} \left( \frac{\partial}{\partial r}  + \frac{l}{r} \right) , \\
\overline{C}^{\uparrow}_{o\to o} &=& r + \frac{1}{2 M T} \left( \frac{\partial}{\partial r}  - \frac{l+1}{r} \right) , \\
\overline{C}^{\downarrow}_{o\to o} &=& r  + \frac{1}{2 M T} \left( \frac{\partial}{\partial r}  + \frac{l}{r} \right) . \label{eq:cdownoo}
\end{eqnarray}
Since the wave function is normalized after application, the normalization of these operators is arbitrary.  In the code, we multiply all of these operators by a factor of $T$ so that they are regular when $T=0$.

\subsection{Width operators in the reduced spherical representation}\label{subsec:gamma_u}
When expressed as operators acting on the reduced wave function, the singlet-to-octet, octet-to-singlet and octet-to-octet width operators take the form
\begin{align}
    \overline{\Gamma}_{s\to o}^\uparrow & =\hat\kappa T^3\frac{l+1}{2l+1}\left[ \left( r -  \frac{N_c \alpha_{\rm s}}{8T} \right)^2-\frac{2l+3}{2MT}+ \frac{\overline{{\cal D}}^2}{(2 M T)^2}+\frac{l+1}{2 M T} \left( \frac{N_c \alpha_{\rm s}}{4T} \right) \frac{1}{r}\right],\label{eq:Gamma_bar_u_s_o}\\ 
    \overline{\Gamma}_{s\to o}^\downarrow &=\hat\kappa T^3\frac{l}{2l+1}\left[\left( r -  \frac{N_c \alpha_{\rm s}}{8T} \right)^2+\frac{2l-1}{2MT}+\frac{\overline{{\cal D}}^2}{(2 M T)^2}-\frac{l}{2 M T} \left( \frac{N_c \alpha_{\rm s}}{4T} \right) \frac{1}{r}\right],\label{eq:Gamma_bar_d_s_o}\\
    \overline{\Gamma}_{o\to s}^{\uparrow} &= \frac{\hat\kappa T^3}{N_c^2-1} \frac{l+1}{2l+1} \left[ \left( r +  \frac{N_c \alpha_{\rm s}}{8T} \right)^2 - \frac{2l+3}{2MT} + \frac{\overline{{\cal D}}^2}{(2 M T)^2} -\frac{l+1}{2 M T} \left( \frac{N_c \alpha_{\rm s}}{4T} \right) \frac{1}{r} \right] ,\label{eq:Gamma_bar_u_o_s}\\
    \overline{\Gamma}_{o\to s}^{\downarrow} &=\frac{\hat\kappa T^3}{N_c^2-1} \frac{l}{2l+1} \left[\left( r +  \frac{N_c \alpha_{\rm s}}{8T} \right)^2+\frac{2l-1}{2MT}+\frac{\overline{{\cal D}}^2}{(2 M T)^2}+\frac{l}{2 M T} \left( \frac{N_c \alpha_{\rm s}}{4T} \right) \frac{1}{r}\right]\,, \label{eq:Gamma_bar_d_o_s}\\
    \overline{\Gamma}_{o\to o}^{\uparrow} &=  \hat\kappa T^3 \frac{N_c^2-4}{2(N_c^2-1)} \frac{l+1}{2l+1} \left[ r^2 - \frac{2l +3}{2MT} + \frac{\overline{{\cal D}}^2}{(2 M T)^2} \right] \,, \label{eq:Gamma_bar_u_o_o}\\
    \overline{\Gamma}_{o\to o}^{\downarrow} &= \hat\kappa T^3 \frac{N_c^2-4}{2(N_c^2-1)}\frac{l}{2l+1}  \left[ r^2 + \frac{2l-1}{2MT} + \frac{\overline{{\cal D}}^2}{(2 M T)^2} \right] \,\label{eq:Gamma_bar_d_o_o},
\end{align}
with
\begin{align}
\overline{{\cal D}}^2 = -\frac{\partial^2}{\partial r^2} + \frac{l (l+1)}{r^2} \, .
\end{align}

\subsection{Effective Hamiltonian in the reduced spherical representation}\label{subsec:h_u}

The effective Hamiltonian for singlet and octet evolution is defined by $H^{\rm eff}_{s,o} = h_{s,o}  + \text{Im}(\Sigma_{s,o})  - i\Gamma_{s,o}/2$ with $\Gamma_s = \sum_{i \in \{\uparrow,\downarrow\}} \Gamma_{s\rightarrow o}^i$ and $\Gamma_o =  \sum_{i \in \{\uparrow,\downarrow\}} (\Gamma_{o\rightarrow s}^i + \Gamma_{o\rightarrow o}^i$).  
When expressed as operators acting on the reduced wave function, the singlet effective Hamiltonian $\overline{H}^{\rm eff}_s$ is given by
\begin{align}
    \text{Re}[\overline{H}^{\rm eff}_s] &=  
    \frac{\overline{\cal D}^2}{M}
    - \frac{C_{f}\, \alpha_{\rm s}}{r} + \frac{\hat\gamma T^3}{2} r^2 +  \frac{\hat\kappa T^2}{4 M} \{r,p_r\} \, , \label{eq:heff1}\\
    \text{Im}[\overline{H}^{\rm eff}_s] &= - \frac{\hat\kappa T^3}{2} \left[ \left( r - \frac{N_c \alpha_{\rm s}}{8T} \right)^2 - \frac{3}{2MT} + \frac{{\overline{\cal D}}^2}{(2MT)^2} + \frac{1}{2MT} \left( \frac{N_c \alpha_{\rm s}}{4 T} \right)  \frac{1}{r}\right] ,
\end{align}
where $p_r = - i \partial_r$.  Similarly, the octet effective Hamiltonian $\overline{H}^{\rm eff}_o$ is given by
\begin{align}
    \text{Re}[\overline{H}^{\rm eff}_o] &=  \frac{\overline{\mathcal{D}}^{2}}{M} + \frac{1}{2N_{c}} \frac{\alpha_{\rm s}}{r} + \frac{N_c^2 - 2}{2(N_c^2 - 1)}\left[ \frac{ \hat\gamma T^3 }{2} r^2 +  \frac{\hat\kappa T^2}{4 M} \{r,p_r\} \right]  , \\
    \text{Im}[\overline{H}^{\rm eff}_o] &= - \frac{\hat\kappa T^3}{2(N_c^2-1)} \left[ \left( r + \frac{N_c \alpha_{\rm s}}{8T} \right)^2 - \frac{3}{2MT} + \frac{{\overline{\cal D}}^2}{(2MT)^2} - \frac{1}{2MT} \left( \frac{N_c \alpha_{\rm s}}{4 T} \right)  \frac{1}{r}\right]  \nonumber \\
&~ \hspace{0.4cm} {-}\frac{\hat\kappa T^3}{4(N_c^2-1)} \left[ r^2  
- \frac{3}{2MT} + \frac{{\overline{\cal D}}^2}{(2MT)^2} 
 \right] ,  \label{eq:heff4}
\end{align}
where $\hat{\kappa} = \kappa / T^3$ and  $\hat{\gamma} = \gamma / T^3$.

\subsection{Description of the algorithm}\label{subsec:qtraj_algorithm}

The NLO quantum trajectories algorithm proceeds as follows:
\begin{enumerate}
    \item Initialize the reduced wave function $u(t_0,r) = u_0(r)$ at $N$ points which are uniformly distributed on $(0,L)$ with $L$ being the box size.  
    The wave function is assumed to obey Dirichlet boundary conditions at the boundaries with $u(t,0)=u(t,L)=0$ at all times.  
    Additionally, we specify the initial color state $c \in \{0,1\}$, with 0 being a color singlet and 1 being a color octet state, and the initial orbital angular momentum quantum number $l \geq 0$.
    \item Normalize the wave function such that $|u|^2 = \langle u | u \rangle$ = 1.\label{step:normalization}
    \item Generate a random number $p_{1}$ uniformly distributed in $[0,1]$.
    \item Using $\overline{H}^{\rm eff}_{s,o}$ specified in eqs.~\eqref{eq:heff1}-\eqref{eq:heff4}, evolve the wave function in time until the jump time $t_j$. A quantum jump is triggered by the norm squared of the wave function falling below $p_{1}$ at time $t_{j}$, i.e., $|u(t_{j},r)|^2 < p_{1}$.
    If the maximum evolution time is reached before a jump is triggered, i.e., $t_{j}>t_{f}$, terminate execution at time $t_{f}$, otherwise, perform a quantum jump at time $t_{j}$ by proceeding to step~\ref{step:quantum_jump}.\label{step:evolution}
    \item Initiate a {\em quantum jump} by generating an additional random number $p_2$ uniformly distributed in $[0,1]$.  
    This number will be used to select the jump operator to apply. \label{step:quantum_jump}
    \item If the color state is singlet ($c=0$):
    \begin{enumerate}
        \item Set the color state to octet, i.e. $c \rightarrow 1$.
        \item Calculate the jump probabilities $P_1$ and $P_2$ from eqs.~(\ref{eq:Gamma_bar_u_s_o}) and~(\ref{eq:Gamma_bar_d_s_o}) as $P_{1(2)}=\braket{\bar{\Gamma}^{\uparrow(\downarrow)}_{s\to o}}/(\braket{\bar{\Gamma}^{\uparrow}_{s\to o}}+\braket{\bar{\Gamma}^{\downarrow}_{s\to o}})$. 
        \item If $p_2 < P_1$, set $l \rightarrow l +1$, else $l \rightarrow l - 1$.
    \end{enumerate}
    Else ($c=1$):
    \begin{enumerate}
        \item Evaluate the transition amplitudes $\overline\Gamma_1 =  \langle \overline\Gamma_{o\to o}^{\downarrow} \rangle$, $\overline\Gamma_2=  \langle \overline\Gamma_{o\to o}^{\uparrow} \rangle$, $\overline\Gamma_3=  \langle \overline\Gamma_{o\to s}^{\downarrow} \rangle$, and $\overline\Gamma_4=  \langle \overline\Gamma_{o\to s}^{\uparrow} \rangle$ using the operators listed in eqs.~(\ref{eq:Gamma_bar_u_o_s})-(\ref{eq:Gamma_bar_d_o_o}) above.  
        From these, obtain the jump probabilities $P_i = \overline\Gamma_i/\sum_{j=1}^{4} \overline\Gamma_j$.  
        \item If $p_2 < P_1$: $l \rightarrow l -1$,\\
         else if $P_1<p_2$ and $p_2<P_1+P_2$: $l \rightarrow l +1$,\\
         else if $P_1+P_2<p_2$ and $p_2<P_1+P_2+P_3$: $l \rightarrow l -1$ and $c\to 0$,\\
         else if $P_1+P_2+P_3<p_2$: $l \rightarrow l +1$ and $c\to 0$.
    \end{enumerate}
\item Based on the choices made above, apply the appropriate collapse/jump operator to the wavefunction $u(r) \rightarrow \overline{C} u(r)$ with $\overline{C} \in \{ \overline{C}_{s \to o}^{\downarrow}, \overline{C}_{s\to o}^{\uparrow}, \overline{C}_{o \to s}^{\downarrow}, \overline{C}_{o\to s}^{\uparrow}, \overline{C}_{o\to o}^{\downarrow}, \overline{C}_{o \to o}^{\uparrow}\}$ given in eqs.~\eqref{eq:cupso}-\eqref{eq:cdownoo} above.  
\item Repeat starting from step~\ref{step:normalization}.
\end{enumerate}

\subsection{Evolution between jumps}\label{subsec:evolution_between_jumps}
In step~\ref{step:evolution} above, we must evolve the wave function forward in time using the complex effective Hamiltonian $\overline{H}^{\rm eff}_{s,o} $.  
Because of the appearance of the anticommutator term $\{r,p_r\}$, it is not easy to straightforwardly apply the same split-step pseudospectral method (Suzuki--Trotter) as was used in ref.~\cite{Omar:2021kra}.  
Instead, here we use the Crank--Nicolson method to update the wave function between quantum jumps.  
With this method, one approximates the infinitesimal time evolution operator as
\be
e^{-i \overline{H}^{\rm eff}_{s,o} \Delta t} \simeq \frac{1 - i\Delta t \overline{H}^{\rm eff}_{s,o}/2}{1 + i\Delta t \overline{H}^{\rm eff}_{s,o}/2} \, ,
\ee
where the Hamiltonian has implicit time dependence.
To proceed, one expresses this as
\be
\left(1 + \frac{i}{2} \Delta t \overline{H}^{\rm eff}_{s,o} \right) \psi(t+\Delta t) = \left(1 - \frac{i}{2} \Delta t \overline{H}^{\rm eff}_{s,o} \right)  \psi(t) \, .
\label{eq:CNupdate1}
\ee
To evolve the wave function one step forward in time, one applies the terms on the right to obtain the ``half updated'' wave function $X \equiv \left(1 - i\Delta t \overline{H}^{\rm eff}_{s,o}/2 \right)  \psi(t)$.  
The final step in the Crank--Nicolson algorithm is to solve
\be
A_{ij}  \psi_j(t+\Delta t) = X_i \, ,
\ee
for $ \psi_j(t+\Delta t)$ where $i,j$ are spatial coordinates and $A_{ij} =  \left(1 +  i \Delta t \overline{H}^{\rm eff}_{s,o}/2 \right)_{ij} $.  Typically $A_{ij}$ is tridiagonal, in which case there are many optimized solvers.  
For the results below, we make use of the optimized sparse matrix solver {\tt spsolve} provided by the open-source Armadillo package \cite{Armadillo}.  
The Armadillo package is also used to handle all matrix multiplications required to compute the action of the effective Hamiltonian, width, and jump operators.

\section{Results}
\label{sec:results}

We consider $5.02$ TeV Pb-Pb collisions with the background temperature evolution given by 3+1D quasiparticle anisotropic hydrodynamics which was tuned to reproduce experimentally observed soft hadron spectra, elliptic flow, and HBT radii  \cite{Alqahtani:2020paa,Alalawi:2021jwn}.  
For this purpose, smooth optical Glauber initial conditions were used, and the resulting initial central temperature was $T_0 = $ 630 MeV at $\tau_0 = 0.25$ fm and the shear viscosity to entropy density ratio was $\eta/s = 0.159$.  
All other transport coefficients, such as the bulk viscosity, were self-consistently determined in terms of $\eta/s$ and the lattice-based equation of state used in the evolution.  
The hydrodynamic evolution used is the same as that used in the prior papers \cite{Brambilla:2020qwo,Brambilla:2021wkt}.

For all results presented herein, we used a one-dimensional lattice with $\texttt{NUM}=2048$ points and $\texttt{L}=40\,\mathrm{GeV}^{-1}$ with a corresponding lattice spacing of $a \simeq 0.0195\,\mathrm{GeV}^{-1}$. The Crank--Nicolson evolution time step was taken to be $\texttt{dt}=0.001\,\mathrm{GeV}^{-1}$.  
This spatiotemporal discretization was shown in prior work to have only small lattice spacing and finite size effects \cite{Omar:2021kra}.  
Finally, we note that the code used to generate all results in this paper is available publicly in the NLO branch of the \qtraj code repository \cite{qtraj-download}.

As emphasized in the introduction, the parameters entering the evolution equations do not need to be fit to experimental data.
These are the heavy quark mass $M$, the strong coupling $\alpha_{\rm s}$, and the transport coefficients $\kappa$ and $\gamma$.
We work in the bottom sector and, as such, take $M = m_{b} = m_{\Upsilon(1S)}/2 = 4.73$ GeV with $m_{\Upsilon(1S)}$ from ref.~\cite{Zyla:2020zbs}.
We compute the Bohr radius $a_0$ from the 1-loop relation
\be
a_{0} = \frac{2}{C_{f} \, \alpha_{\rm s}(1/a_{0}) \, m_{b}},
\ee
and evaluate $\alpha_{\rm s}$ at the inverse of the Bohr radius using the 1-loop running with $N_{f}=3$ flavors and $\Lambda_{\overline{\rm MS}}^{N_f=3}=332$ MeV~\cite{Zyla:2020zbs} obtaining $\alpha_{\rm s}(1/a_0) = 0.468$.

The transport coefficients $\kappa$ and $\gamma$ are fixed from direct and indirect lattice measurements, respectively.
In ref.~\cite{Brambilla:2020siz}, the heavy quark momentum diffusion coefficient \cite{CasalderreySolana:2006rq,CaronHuot:2007gq} was measured directly in a quenched lattice simulation over a wide range of temperatures. 
Here we identify $\kappa(T)$ with this lattice determination of the quark momentum diffusion coefficient, while we postpone a discussion on this identification and its associated uncertainty to appendix~\ref{app:kappa}.
We perform \texttt{QTraj} simulations using three parametrizations of the dimensionless quantity $\hat{\kappa}(T)=\kappa / T^{3}$ which we denote $\hat{\kappa}_{L}(T)$, $\hat{\kappa}_{C}(T)$, and $\hat{\kappa}_{U}(T)$ and which correspond to the lower, central, and upper ``fit'' curves, respectively, of fig.~13 of ref.~\cite{Brambilla:2020siz}; we use this variation as an estimate of our systematic uncertainty due to $\kappa$.
As no direct lattice measurements of $\gamma$ currently exist, we make use of an indirect estimate from unquenched lattice simulations.
At zeroth order in the $E/T$ expansion, projection of the in-medium heavy quarkonium self energy onto the $1S$ state yields the relation $\delta M(1S) = (3/2)a_{0}^{2} \gamma$, where $\delta M(1S)$ is the in-medium mass shift of the $1S$ state, cf. eq.~(78) of ref.~\cite{Brambilla:2017zei} and ref.~\cite{Brambilla:2019tpt}.
From this relation, $\gamma$ is accessible from unquenched lattice measurements of $\delta M(1S)$.
In ref.~\cite{Brambilla:2019tpt}, unquenched lattice measurements of  $\delta M[\Upsilon(1S)]$ from refs.~\cite{Kim:2018yhk,Aarts:2011sm} were used to place bounds on $\hat{\gamma}(T) = \gamma / T^{3}$ of approximately  $-3.5 \leq \hat\gamma \leq 0$.
We perform simulations using $\hat{\gamma}$ in this range and use this as an estimation of our systematic uncertainty due to $\gamma$.
We note that more recent lattice studies \cite{Larsen:2019bwy,Shi:2021qri} seem to favor $\delta M(\Upsilon(1S)) \simeq 0$ and thus $\hat{\gamma} \simeq 0$.

To initialize the simulation, we assume that at $\tau = 0$ fm the bottomonium wave function is in a singlet state with a smeared delta function initial condition.\footnote{One can also consider octet initialization; however, we find that the off-diagonal octet-singlet overlap is negligible, similar to the finding at leading order \cite{Brambilla:2020qwo}.}   In practice, the initial reduced wave function is given by a Gaussian delta function multiplied by a power of $r$ appropriate for the initial angular momentum state $l$, i.e.,
\begin{equation}
u_{\ell}(t_0) \propto r^{l+1} e^{-r^{2}/(ca_{0})^2},
\end{equation} 
with $u$ normalized to one when summed over the entire (one-dimensional) lattice volume. 
We set $c=0.2$ noting that observables do not appear to show significant dependence on this parameter for $c$ below this value (cf. fig.~6 of ref.~\cite{Omar:2021kra}) while the computational cost increases dramatically as $c$ is decreased.
For the results presented in the main body of the text, we evolve the initial wave function using the vacuum potential ($\hat\kappa=\hat\gamma=0$) from $\tau = 0$~fm to $\tau_{\rm med}$ = 0.6~fm at which time we turn on the medium interactions.  
This is the same medium initialization time scale as used in refs.~\cite{Brambilla:2020qwo,Brambilla:2021wkt}.
To assess the dependence on this assumption, in appendix~\ref{app:taumed}, we present results obtained with $\tau_{\rm med}$ = 0.25 fm, which corresponds to the earliest time for which the anisotropic hydrodynamic simulation results are available.

\subsection{Comparison of LO and NLO singlet-octet widths}
\label{sec:widthcomp}

\begin{figure}[ht]
\begin{center}
\includegraphics[width=0.8\linewidth]{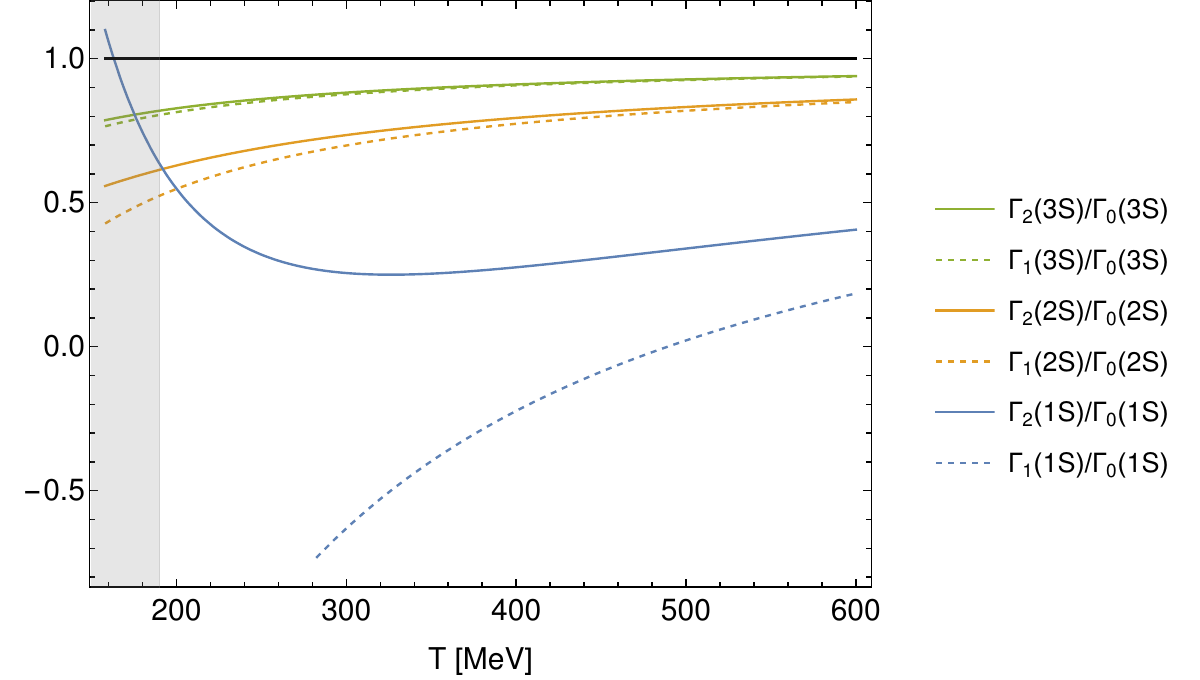}
\end{center}
\caption{
The singlet to octet widths of the 1S, 2S, and 3S states including $E/T$ corrections.
$\Gamma_{m}(nS)$ represents the singlet to octet width of the state $nS$ including terms up to order $(E/T)^{m}$.
The black line at unity represents the limit of perfect convergence of the $E/T$ expansion, i.e., $E/T \to 0$.
The plot spans from $T=158$ MeV to $T=600$ MeV; the gray shaded area on the left represents the temperature region $T<T_{f}=190$ MeV not included in the phenomenological results presented in this work.
}
\label{fig:widths}
\end{figure}

In this subsection, we analyze the effect of the inclusion of higher order terms in the $E/T$ expansion. Projecting the singlet to octet width $\Gamma_{s\to o}$ given in eq.~(\ref{eq:Gamma_s_to_o}) onto $1S$, $2S$, and $3S$ states, we find
\begin{align}
    \langle 1S | \Gamma_{s\to o} | 1S \rangle &= 3 a_{0}^{2} \kappa \left\{
    1 - \frac{2N_{c}^{2}-1}{2(N_{c}^{2}-1)}\frac{E}{T}
    + \frac{(2N_{c}^{2}-1)^{2}}{12(N_{c}^{2}-1)^{2}}\left(\frac{E}{T}\right)^{2}\right\},\\
    \langle 2S | \Gamma_{s\to o} | 2S \rangle &= 42 a_{0}^{2} \kappa \left\{
    1 - \frac{5N_{c}^{2}-1}{28(N_{c}^{2}-1)}\frac{E}{T}
    + \frac{7N_{c}^{4}-4N_{c}^{2}+1}{672(N_{c}^{2}-1)^{2}}\left(\frac{E}{T}\right)^{2}\right\},\\
    \langle 3S | \Gamma_{s\to o} | 3S \rangle &= 207 a_{0}^{2} \kappa \left\{
    1 - \frac{10N_{c}^{2}-1}{138(N_{c}^{2}-1)}\frac{E}{T}
    + \frac{12N_{c}^{4}-4N_{c}^{2}+1}{7452(N_{c}^{2}-1)^{2}}\left(\frac{E}{T}\right)^{2}\right\},
\end{align}
where $E = 1/(Ma_0^2)$ is the magnitude of the Coulombic binding energy; the above expressions make explicit the $E/T$ expansion.
We plot the various contributions to the bottomonium widths in fig.~\ref{fig:widths}.
Reading the plot from left to right, we observe the expansion to converge for high $T$ and, comparing the solid to dotted curves, upon inclusion of additional terms in the $E/T$ expansion.
Furthermore, comparing the blue to orange to green curves, we observe better convergence properties for states of higher principal quantum due to their lower Coulombic binding energy.

\subsection{Comparisons of LO and NLO survival probabilities with complex \texorpdfstring{$H_{\rm eff}$}{Heff}}

In this subsection, we investigate the effect of the NLO evolution terms on the survival probabilities of the $\Upsilon(1S)$ state.
To model the medium evolution, we follow the procedure of ref.~\cite{Brambilla:2020qwo} and sample a large number of physical bottomonium trajectories through the QGP recording the temperature along each trajectory.  
We bin the events in centrality (0-100\%) and average over the physical trajectories in a given centrality bin to obtain a trajectory-averaged temperature evolution in each bin; plots of the temperature evolution in each centrality bin considered can be found in fig.~4 of ref.~\cite{Brambilla:2020qwo}.

\begin{figure}[ht]
\begin{center}
\begin{subfigure}{0.49\textwidth}
    \includegraphics[width=1\textwidth]{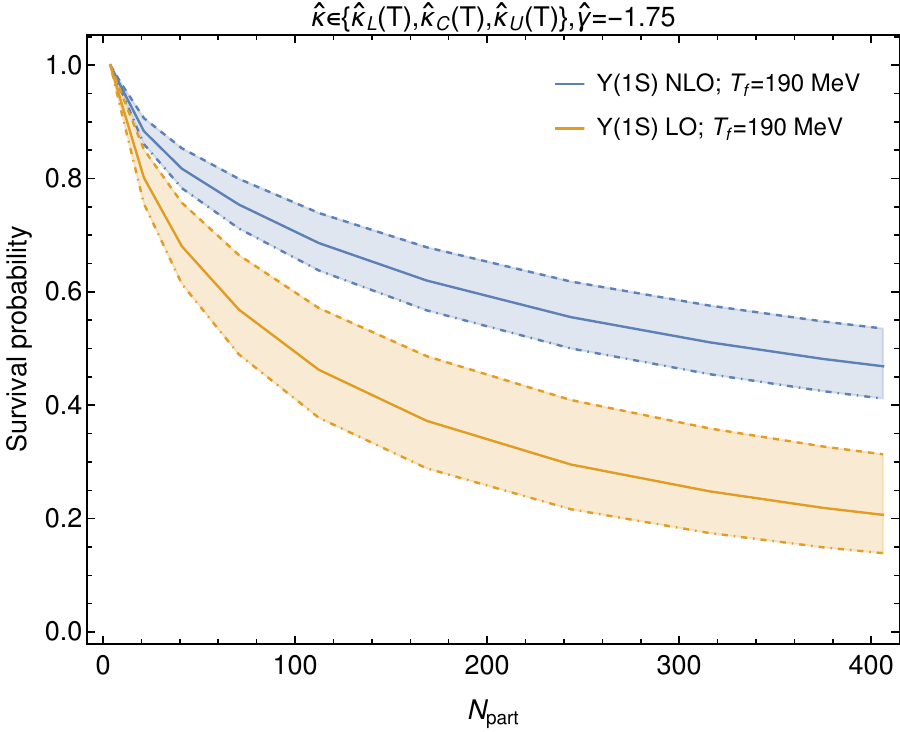}
\end{subfigure}
\begin{subfigure}{0.49\textwidth}
    \includegraphics[width=1\textwidth]{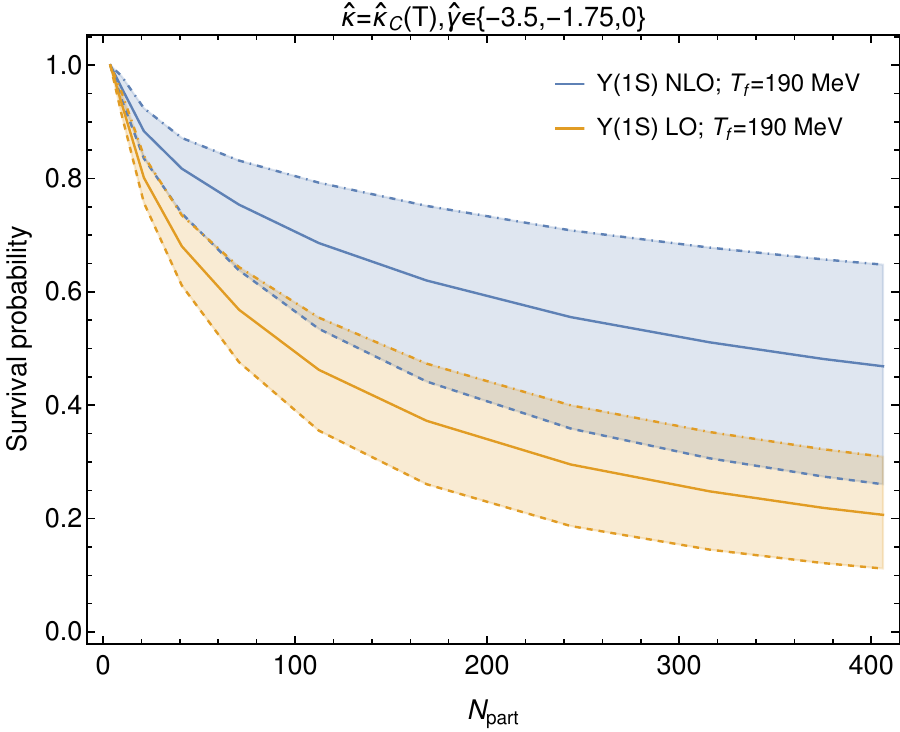}
\end{subfigure}
\begin{subfigure}{0.49\textwidth}
    \includegraphics[width=1\textwidth]{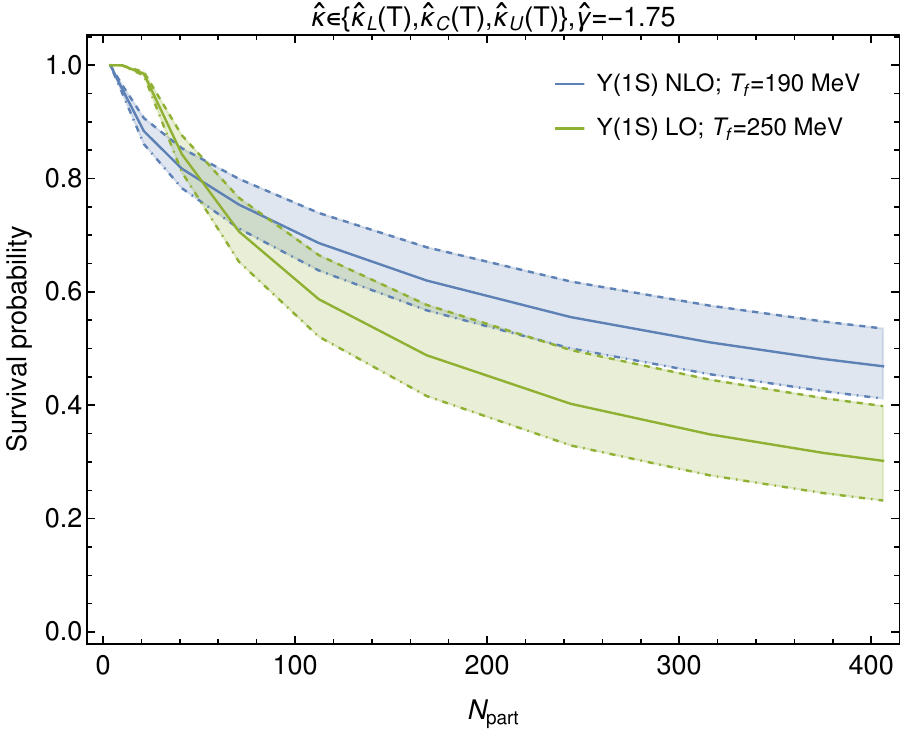}
\end{subfigure}
\begin{subfigure}{0.49\textwidth}
    \includegraphics[width=1\textwidth]{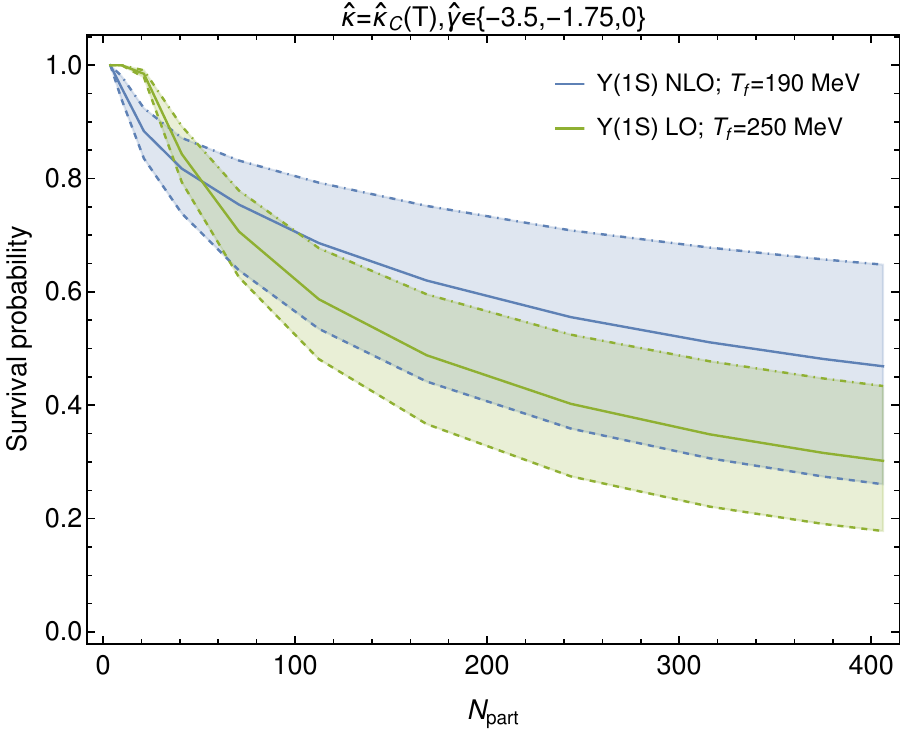}
\end{subfigure}
\end{center}
\caption{
The survival probability as a function of $N_{\text{part}}$ of the $1S$ state calculated using $H_{\rm eff}$ evolution without jumps.
In the top row, we compare results obtained using LO and NLO evolution both with $T_{f}=190$ MeV to highlight the effect of the inclusion of higher order terms in the $E/T$ expansion.
In the bottom row, we compare our current state of the art NLO evolution with $T_{f}=190$ MeV to LO evolution with $T_{f}=250$ MeV as used in previous works.
The bands indicate variation with respect to $\hat{\kappa}(T)$ (left) and $\hat{\gamma}$ (right).
The central curves represent the central values of $\hat{\kappa}(T)$ and $\hat{\gamma}$, and the dashed and dot-dashed lines represent the lower and upper values, respectively, of $\hat{\kappa}(T)$ and $\hat{\gamma}$.
}
\label{fig:heff}
\end{figure}

Using this trajectory averaged temperature evolution, we perform \texttt{QTraj} simulations including only the LO and the NLO terms of the evolution.
For the LO evolution, we run simulations down to $T_{f}=190$ MeV and $T_{f}=250$ MeV; for the NLO evolution, we use $T_{f}=190$ MeV.
We plot these results in fig.~\ref{fig:heff}.
Comparing the LO and NLO results both with $T_{f}=190$ MeV, we observe the former to fall below the latter for all but the most peripheral events where very little hydrodynamic evolution takes place.
Comparing the LO results with $T_{f}=250$ MeV to the NLO results with $T_{f}=190$ MeV, we observe the NLO results to fall above the LO results in central collisions and below them in peripheral collisions.
There are two competing trends at work here: that of the NLO terms to reduce suppression and that of lower $T_{f}$ to increase it.
The former dominates in central collisions with greater total time spent traversing the QGP while the latter dominates in peripheral collisions for which a high value of $T_{f}$ can lead to no hydrodynamic evolution due to low initial temperatures.

We note that in the upper panels of fig.~\ref{fig:heff}, the LO results are evolved beyond their range of validity, i.e., to regions in which the strict $E \ll T$ hierarchy in which the LO order evolution equations were derived no longer holds.
This leads to larger differences between the LO and NLO results than in the lower panels of fig.~\ref{fig:heff} in which both the LO and NLO results are evolved only within their range of validity.
Furthermore, in the lower panels, in the regions where hydrodynamic evolution takes place, we observe consistency between the LO and NLO results up to the systematic uncertainties due to the transport coefficients.
We interpret this consistency as signaling the stability and predictivity of the results when simulating within the temperature range in which the hierarchy of scales in which the evolution equations are derived is valid.
We have also verified that the relative difference between the LO and NLO survival probabilities of the $1S$ state with Bjorken evolution from $T_{0}=425$ MeV remains $\lesssim 50\%$ down to $T_{f}=190$ MeV; this $50\%$ relative difference may be taken as signalling, from a phenomenological perspective, the breakdown of the $E/T$ expansion.
We note, furthermore, that this difference is largest for the $1S$ state.
We consider this consistent with the effect on, e.g., the in-medium width of the $\Upsilon(2S)$ displayed in fig.~\ref{fig:widths}.

\begin{figure}[ht]
\begin{center}
\includegraphics[width=\linewidth]{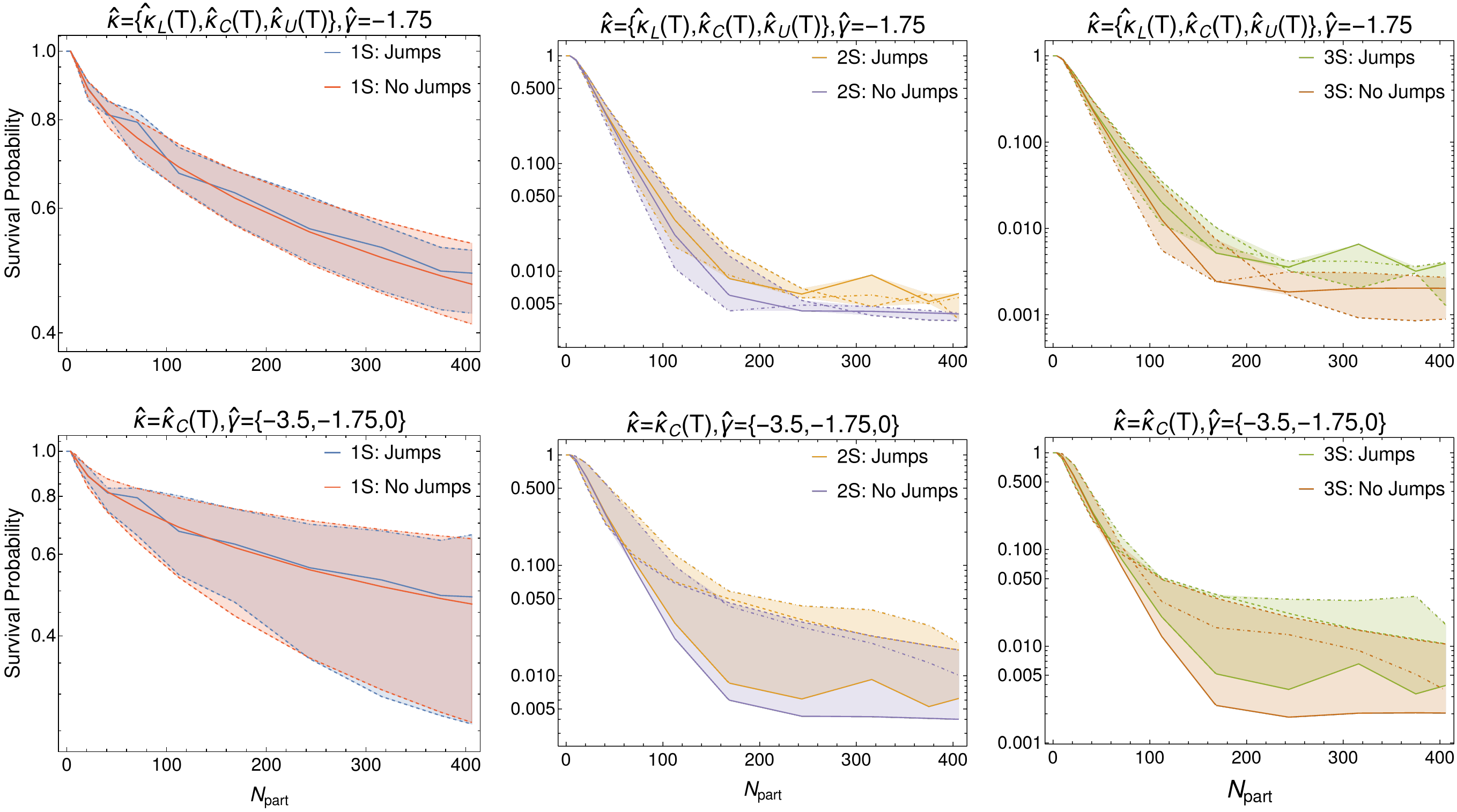}
\end{center}
\caption{Jumps vs. no jumps.  
Top row is $\hat{\kappa}$ variation; bottom row is $\hat{\gamma}$ variation.  
Solid, dashed, and dot-dashed curves represent the central, lower, and upper values, respectively, of $\hat{\kappa}$ and $\hat{\gamma}$.}
\label{fig:raavsnpart-jumpnojump}
\end{figure}

\subsection{Assessing the effects of quantum jumps on the survival probabilities}
\label{subsec:jumpnojump}

In this subsection, we assess the effects of quantum jumps on our NLO results.
We make use of the same trajectory-averaged temperature profile as in the previous subsection.
In fig.~\ref{fig:raavsnpart-jumpnojump}, we present results for the survival probability of the $\Upsilon(1S)$, $\Upsilon(2S)$, and $\Upsilon(3S)$ states as functions of centrality.
For the results including full evolution with stochastically sampled jumps, we report only the central (mean) values and note that the statistical uncertainties are sub-leading when compared to the uncertainty resulting from $\kappa$ and $\gamma$ variation.
For identical parameters, we observe the jumps to increase the yield of the states.
The relative effect appears to become more pronounced for the excited states.
We note, however, that the low yields of the excited states make this difference small in absolute terms.
As in the leading order case explored in refs.~\cite{Brambilla:2020qwo,Brambilla:2021wkt}, the lower probability of octet to singlet transitions compared to octet to octet transitions combined with the repulsive octet potential driving the state to large radii with little overlap with the ground and first excited states makes the effect of the stochastic jumps much smaller than the theoretical uncertainties stemming from $\kappa$ and $\gamma$ variation.
For $R_{AA}$, the evolution with $H_{\text{eff}}$ provides a reasonable estimation at a significantly reduced computational cost.

\subsection{NLO \texorpdfstring{$H_{\rm eff}$}{Heff} results for \texorpdfstring{$R_{AA}$}{RAA} with sampled physical trajectories}

In this subsection we present results of $H_{\rm eff}$ evolution using an ensemble of 80,000 sampled physical trajectories for each set of values for $\hat\kappa(T)$ and $\hat\gamma$.  Unlike the results presented in the previous two subsections, here we compute the survival probability on a trajectory-by-trajectory basis and average the resulting survival probabilities rather than using a single trajectory-averaged temperature evolution in each centrality bin.  Using individually sampled physical trajectories allows us to more faithfully include the physics of states produced at a large distance from the center of the overlap region (the corona effect).  This is particularly important for quantitatively understanding excited state survival probabilities,  
since the corona region contributes a large fraction of surviving states (see, e.g., fig.~3 of ref.~\cite{Islam:2020bnp}).  

\subsubsection*{Excited state feed down}

The Lindblad equation, which the \texttt{QTraj} code solves, gives the probability of a heavy quark-antiquark pair exiting the QGP with specific quantum numbers.
In order to compare to experimental measurements of $R_{AA}$, this raw survival probability data must be post processed to account for the probability of a state decaying after exiting the QGP but before reaching the detector.
This effect is denoted \textit{feed down} and can be accounted for by using the feed down matrix $F$ defined as relating the experimental and direct production cross sections, $\vec{\sigma}_{\text{exp}} = F \vec{\sigma}_{\text{direct}}$.
The cross section vectors correspond to the states considered, while $F$ is a matrix the values of which are fixed by the branching fractions of the excited states.
In our analysis, the states considered are $\vec{\sigma} = \{ \Upsilon(1S),\,$ $\Upsilon(2S),\,$ $\chi_{b0}(1P),\,$ $\chi_{b1}(1P),\,$ $\chi_{b2}(1P),\,$ $\Upsilon(3S),\,$ $\chi_{b0}(2P),\,$ $\chi_{b1}(2P),\,$ $\chi_{b2}(2P)\}$.
The entries of $F$ are
\begin{equation}
	F_{ij} = \left\{ \begin{matrix}
		\text{branching fraction $j$ to $i$}, & \text{for } i < j, \\
		1, & \text{for } i = j, \\
		0, & \text{for } i > j,
		\end{matrix} \right.
\end{equation}
where the branching fractions are taken from the Particle Data Group ~\cite{Zyla:2020zbs} (see also eq.~(6.4) of ref.~\cite{Brambilla:2020qwo}).

Finally, the nuclear modification factor $R_{AA}^i$ for state $i$ is given by
\be
R^{i}_{AA}(c,p_T,\phi) = \frac{\left(F \cdot S(c,p_T,\phi) \cdot \vec{\sigma}_{\text{direct}}\right)^{i}}{\vec{\sigma}_{\text{exp}}^{i}} \, ,
\label{eq:feeddown}
\ee
where $S(c,p_T,\phi)$ represents the survival probabilities computed from the \qtraj code in the form of a diagonal matrix; $c$ labels the centrality class, $p_T$ the transverse momentum, and $\phi$ the azimuthal angle.
We use the integrated experimental cross sections 
$\vec{\sigma}_{\text{exp}}=\{57.6$, 19, 3.72, 13.69, 16.1, 6.8, 3.27, 12.0, $14.15\}$ nb,
computed from the measurements of refs.~\cite{Sirunyan:2018nsz,Aaij:2014caa} as explained in sec.~6.4 of ref.~\cite{Brambilla:2020qwo}.

\begin{figure}[ht]
\begin{center}
\includegraphics[width=0.48\linewidth]{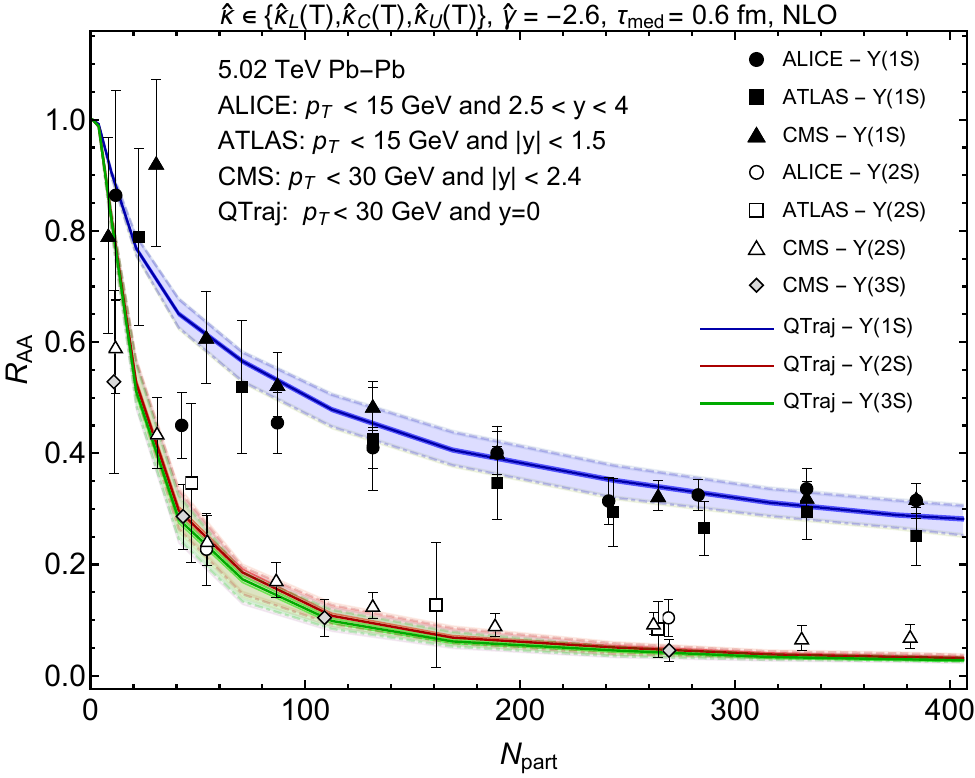} \hspace{3mm}
\includegraphics[width=0.48\linewidth]{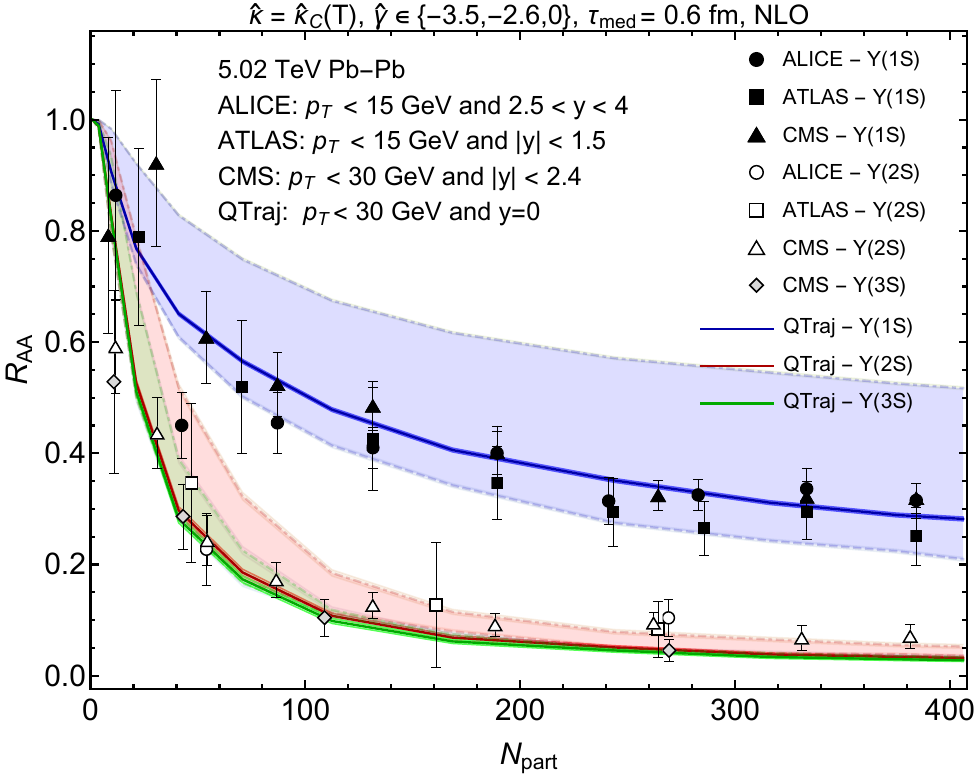}
\end{center}
\caption{
$R_{AA}$ for the $\Upsilon(1S)$, $\Upsilon(2S)$, and $\Upsilon(3S)$ as a function of $N_{\rm part}$.  The left panel shows variation of $\hat\kappa \in \{ \kappa_L(T), \kappa_C(T), \kappa_U(T) \}$ and the right panel shows variation of $\hat\gamma$ in the range $-3.5 \leq \hat\gamma \leq 0$.  In both panels, the solid line corresponds to $\hat\kappa = \hat\kappa_C(T)$ and the best fit value of $\hat\gamma = -2.6$.  The experimental measurements shown are from the ALICE~\cite{Acharya:2020kls}, ATLAS~\cite{ATLAS5TeV}, and CMS~\cite{Sirunyan:2018nsz,CMSupsilonQM2022} collaborations.
}
\label{fig:raavsnpart1}
\end{figure}

\begin{figure}[ht]
\begin{center}
\includegraphics[width=0.48\linewidth]{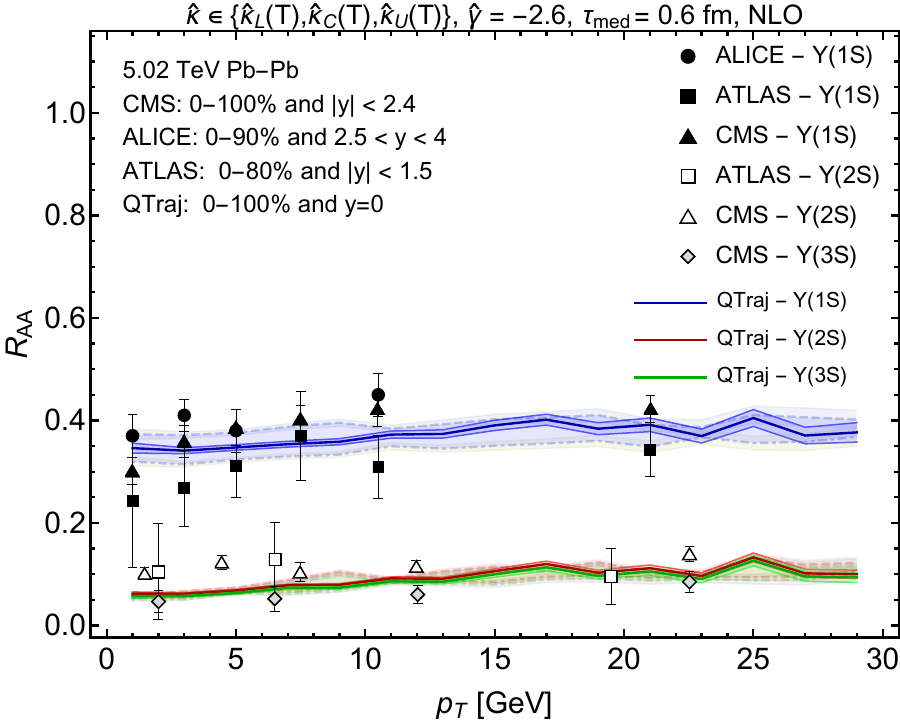} \hspace{3mm}
\includegraphics[width=0.48\linewidth]{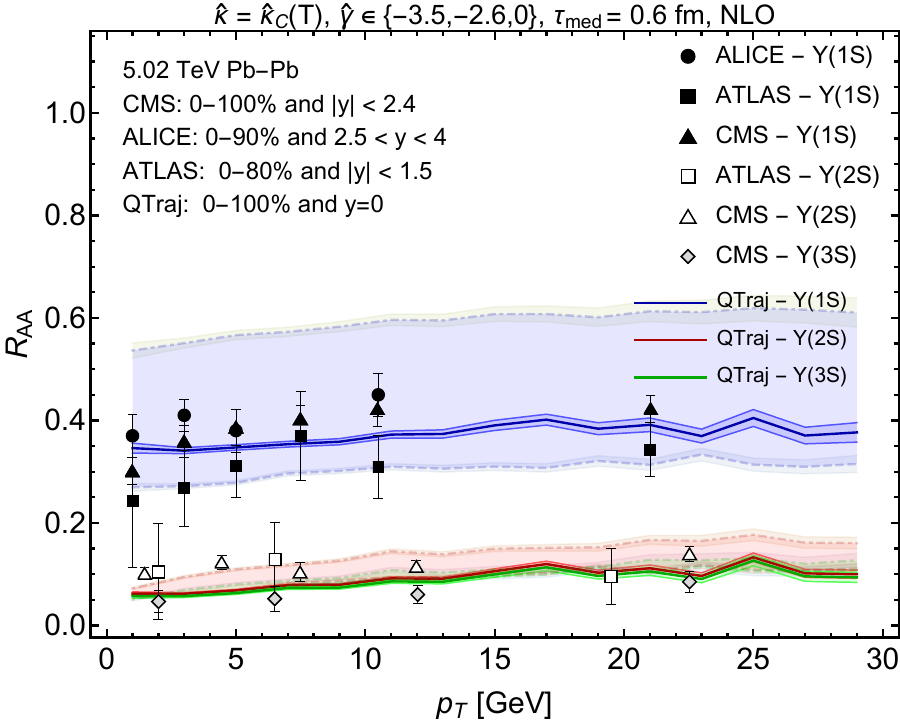}
\end{center}
\caption{
$R_{AA}$ for the $\Upsilon(1S)$, $\Upsilon(2S)$, and $\Upsilon(3S)$ as a function of $p_T$.  The bands and experimental data sources are the same as fig.~\ref{fig:raavsnpart1}.
}
\label{fig:raavspt1}
\end{figure}

\subsubsection*{NLO predictions for $R_{AA}$ using \texorpdfstring{$H_{\rm eff}$}{Heff} evolution}

We now turn to our results for $R_{AA}$ of the 1S, 2S, and 3S states.  
In figs.~\ref{fig:raavsnpart1} and \ref{fig:raavspt1}, we present our NLO predictions for $R_{AA}$ as a function of $N_{\rm part}$ and $p_T$, respectively.  
The results shown in both figures include the effect of excited state feed down using the method described in the previous subsection, and, as in previous sections, we take the decoupling temperature to be $T_f$ = 190 MeV and $\tau_{\rm med{}} = 0.6$ fm.\footnote{We consider the case of $T_f$ = 190 MeV and $\tau_{\rm med}$ = 0.25 fm in appendix~\ref{app:taumed}.} 
For the results presented in figs.~\ref{fig:raavsnpart1} and \ref{fig:raavspt1}, we do not include the effect of quantum jumps due to the high numerical demand when including them; quantum jumps require approximately 50-100x more computational resources than using only the complex effective Hamiltonian evolution.  
Based on the results presented and discussed in  sec.~\ref{subsec:jumpnojump}, we expect that using $H_{\rm eff}$ without quantum jumps is a reliable approximation for the suppression of the $\Upsilon(1S)$; however, we expect this approximation to over-predict the amount of excited state suppression compared to the complete solution of the Lindblad equation. 

In the left panel of fig.~\ref{fig:raavsnpart1}, we show the variation of $\hat\kappa$ in the set $\hat\kappa \in \{ \hat\kappa_L(T), \hat\kappa_C(T), \hat\kappa_U(T) \}$ while holding $\hat\gamma = -2.6$.  
This value of $\hat\gamma$ was chosen as to best reproduce the collected $\Upsilon(1S)$ suppression data.  
In the right panel of fig.~\ref{fig:raavsnpart1}, we show the variation of $\hat\gamma$ in the range $-3.5 \leq \hat\gamma \leq 0$.  
As in the left panel, the solid line corresponds to $\hat\gamma = -2.6$.  
In both the left and right panels, the experimental data presented are from the ALICE~\cite{Acharya:2020kls}, ATLAS~\cite{ATLAS5TeV}, and CMS~\cite{Sirunyan:2018nsz,CMSupsilonQM2022} collaborations. 

Compared to the LO results presented in ref.~\cite{Brambilla:2021wkt}, we find that the variation of $R_{AA}$ with $\hat\kappa$ is reduced, while the variation with $\hat\gamma$ is increased.  
The former is due to the fact that, for fixed $\hat\kappa$, going from LO to NLO reduces the singlet to octet decay width.  
As a result, varying $\hat\kappa$ over a fixed range results in a smaller variation of the dimensionful singlet to octet decay width.  
Regarding the $\hat\gamma$ variation, it is not entirely clear to us why the $R_{AA}$ variation becomes larger when going from LO to NLO, however, this indicates that there is a higher degree of dynamical quantum mixing between states at NLO during the real-time evolution.  
Apart from these changes in the sensitivity to $\hat\kappa$ and $\hat\gamma$, we find that our NLO predictions with $T_f = $190 MeV and our LO predictions with $T_f = 250$ MeV presented in ref.~\cite{Brambilla:2021wkt} are in good agreement, with the NLO result being in slightly better agreement with data, particularly at low $N_{\rm part}$.  
In both figs.~\ref{fig:raavsnpart1} and \ref{fig:raavspt1}, we find excellent agreement between our NLO $H_{\rm eff}$ predictions and the experimental data for the $R_{AA}[1S]$, however, for the 2S and 3S excited states, the $H_{\rm eff}$ predictions are somewhat lower than the experimental results particularly for the most central collisions.  
As demonstrated in fig.~\ref{fig:raavspt1}, when integrated over centrality, the $p_T$-dependence of $R_{AA}$ is better reproduced.

Looking forward, the underpredictions of $R_{AA}[2S]$ and $R_{AA}[3S]$ indicate that it is perhaps necessary to include the effect of quantum jumps on the excited bottomonium states in order to draw firm conclusions.  
As discussed in sec.~\ref{subsec:jumpnojump}, the inclusion of quantum jumps does not strongly affect the  $1S$ survival probability, whereas it increases both the $2S$ and $3S$ survival probabilities relative to the $H_{\rm eff}$ (no jump) evolution.  
We plan to present results including the effects of quantum jumps in a followup paper.

\section{Conclusions and outlook}
\label{sec:conclusions}

In the paper, we have studied bottomonium suppression by combining pNRQCD with an open quantum system framework.  
Working in the regime $Mv \gg T \gg E$, we have gone beyond prior studies \cite{Brambilla:2020qwo,Brambilla:2021wkt} by deriving and numerically solving a Lindblad-type evolution equation that is accurate to NLO in $E/T$.  
The goal of extending prior calculations to NLO in $E/T$ was (1) to allow the pNRQCD+OQS framework to be applied at lower temperatures than is permitted by a LO truncation and (2) to include terms which are necessary to describe the approach of the system to equilibrium at late times.  

On the first point, we have shown that when going from LO to NLO there is a sizable correction to the singlet decay width for $1S$ bottomonium and smaller corrections for the $2S$ and $3S$ bottomonium states.  
The size of the corrections for the $1S$, in particular, stems from the fact that for the $1S$ state $E/T$ is not small at phenomenologically relevant temperatures.  
Based on our findings in sec.~\ref{sec:widthcomp}, the inclusion of NLO corrections has allowed us to extend the pNRQCD+OQS treatment down to temperatures in the vicinity of the QCD phase transition; in practice, we have lowered the decoupling temperature from $T_f$ = 250 MeV at LO to $T_f$ = 190 MeV at NLO.
We have found that including the NLO corrections improves the description of $R_{AA}[1S]$ when compared to experimental data, particularly at low values of $N_{\rm part}$, which correspond to peripheral and semi-peripheral collisions where the plasma temperature generated is low.
In terms of sensitivity to the parameters, at NLO we have found that the pNRQCD+OQS predictions for $R_{AA}[1S]$ are less sensitive to $\kappa$ since the magnitude of the decay width decreases when going from LO to NLO; however, the sensitivity to $\gamma$ is slightly increased even though the NLO corrections do not affect the mass shift.\footnote{
    We note that $\left \langle \psi \left| \left\{ r_{i}, p_{i} \right\} \right| \psi \right \rangle = 0 $ for $\psi^{*}(r) = \psi(r)$ as is the case for eigenstates of the vacuum Hamiltonian in the reduced spherical representation.
    }
As a result, the increased sensitivity must be due to increased quantum state mixing during the real-time evolution.

Regarding the question of thermalization, qualitatively Abelian computations lead us to expect that the inclusion of NLO corrections in $E/T$ improves the approach to thermal equilibrium of the master equation \cite{Blaizot:2017ypk}. 
Checking that the inclusion of the NLO corrections leads to a thermal ensemble of bottomonium states at late times is beyond the scope of this work since this requires the solution of the Lindblad equation including jumps to be carried out on a much longer time scale than presented here.
This is due to the rather slow rate of equilibration for bottomonium states as seen in prior studies~\cite{Akamatsu:2018xim,Miura:2019ssi}.  
However, we do observe that NLO corrections make the $\Upsilon(1S)$ population more stable. 
This is compatible with the fact that finite energy effects included at NLO make transitions that liberate energy more likely than those that absorb energy.
We plan to study thermalization and report our results in a dedicated forthcoming publication \cite{thermalization}.

\acknowledgments{
	N.B., P.V.G. and A.V. acknowledge support by the DFG cluster of excellence ORIGINS funded by the Deutsche Forschungsgemeinschaft under Germany's Excellence Strategy - EXC-2094-390783311.
	This work has also been supported by European Research Council project ERC-2018-ADG-835105 YoctoLHC; by Maria de Maetzu excellence program under project CEX2020-001035-M; by Spanish Research State Agency under project PID2020-119632GB-I00; and by Xunta de Galicia (Centro singular de investigaci\'on de Galicia accreditation 2019-2022), by European Union ERDF.
	M.S., A.I., and A.T. were supported by the U.S. Department of Energy, Office of Science, Office of Nuclear Physics Award No.~DE-SC0013470.
	M.S. also thanks the Ohio Supercomputer Center for support under the auspices of Project No.~PGS0253.  
}

\appendix

\section{Correlator identities and transport coefficients}\label{app:correlatoridentities}

In this appendix, we derive the correlator identities necessary to write the evolution equations at NLO in $E/T$ in terms of the transport coefficients $\kappa$ and $\gamma$.
The transport coefficient $\kappa$ is defined in terms of the chromoelectric correlator as
\begin{equation}\label{eq:kappadefinition}
	\kappa = \frac{g^{2}}{6 N_{c}}\text{Re}\int^{+\infty}_{-\infty} \text{d}t \, \Big \langle T \tilde{E}^{a}_i(t,\vec{0})\tilde{E}^{a}_i(0,\vec{0}) \Big \rangle = \frac{g^{2}}{6 N_{c}}\int^{+\infty}_{0} \text{d}t \, \Big \langle \left\{ \tilde{E}^{a}_i(t,\vec{0}),\tilde{E}^{a}_i(0,\vec{0}) \right\} \Big \rangle.
\end{equation}
In ref.~\cite{Brambilla:2017zei}, the dispersive counterpart of $\kappa$, denoted $\gamma$, was first defined as
\begin{equation}
	\gamma = \frac{g^{2}}{6 N_{c}}\text{Im}\int^{+\infty}_{-\infty} \text{d}t\, \Big \langle T \tilde{E}^{a}_i(t,\vec{0})\tilde{E}^{a}_i(0,\vec{0}) \Big \rangle = -i \frac{g^{2}}{6 N_{c}}\int^{+\infty}_{0} \text{d}t \, \Big \langle \left[ \tilde{E}^{a}_i(t,\vec{0}),\tilde{E}^{a}_i(0,\vec{0}) \right] \Big \rangle,
\end{equation}
where the $T$ appearing in the correlators expressions represents the time ordering operator.

Integrals over in-medium correlators  at LO in $E/T$ 
can be written in terms of these two transport coefficients, since 
\begin{equation}
	\frac{g^{2}}{6 N_{c}}\int_{0}^{\infty}\text{d}t \, \Big \langle \tilde{E}^{a}_i(t,\vec{0})\tilde{E}^{a}_i(0,\vec{0}) \Big \rangle = \frac{1}{2}\left(\kappa+i\gamma\right).
\end{equation}
For the NLO terms, we have to calculate also integrals 
with an additional factor of $t$ in the integrand.
In order to compute them, 
we start by approximating~\cite{Akamatsu:2012vt,Blaizot:2017ypk}\footnote{
This amounts at expanding the chromoelectric correlator around its instantaneous limit.
}
\begin{equation}\label{eq:postulate}
	i\int_{0}^{\infty}\text{d}t \, t \Big \langle \tilde{E}^{a}_i(t,\vec{0})\tilde{E}^{a}_i(0,\vec{0}) \Big \rangle \simeq
	\frac{1}{2}\left.\frac{\text{d}D^{>}(\omega)}{\text{d}\omega}\right|_{\omega=0},
\end{equation}
where $D^{>}(\omega)$ is the Fourier transform of the chromoelectric correlator,
\begin{equation}
	\int_{-\infty}^{\infty}\text{d}t \, e^{i\omega t} \Big \langle \tilde{E}^{a}_i(t,\vec{0})\tilde{E}^{a}_i(0,\vec{0}) \Big \rangle = D^{>}(\omega).
\end{equation}

Then we define the finite temperature correlators
\begin{alignat}{2}
	D^{>}(t,t') &= \Big \langle \tilde{E}^{a}_i(t,\vec{0})\tilde{E}^{a}_i(t',\vec{0}) \Big \rangle &&= \frac{1}{Z(T)} \text{Tr}\left( \tilde{E}^{a}_i(t,\vec{0}) \tilde{E}^{a}_i(t',\vec{0}) e^{-H/T} 	\right), \\
	D^{<}(t,t') &= \Big \langle \tilde{E}^{a}_i(t',\vec{0})\tilde{E}^{a}_i(t,\vec{0}) \Big \rangle &&= \frac{1}{Z(T)} \text{Tr}\left( \tilde{E}^{a}_i(t',\vec{0}) \tilde{E}^{a}_i(t,\vec{0}) e^{-H/T}
	\right),
\end{alignat}
where $H$ is the Hamiltonian, and $Z(T)$ is the partition function defined by
\begin{equation}
	Z(T) = \text{Tr}\left( e^{-H/T} \right).
\end{equation}
That $e^{-H/T}$ is the time evolution operator in imaginary time, i.e., $e^{-H/T}E^{a}_i(t,\vec{0})e^{H/T}=E^{a}_i(t+i/T,\vec{0})$, implies the Kubo--Martin--Schwinger relation
\begin{equation}
		D^{>}(t,t')=
		D^{<}(t+i/T,t').
\end{equation}
Assuming $D^{>}(t,t') = D^>(t-t')$,   $D^{<}(t,t') = D^<(t-t')$, and setting $t'=0$, we find 
\begin{eqnarray}
		&& D^{>}(t)  = D^{<}(-t) = D^{<}(t+i/T) \nonumber\\
    \Rightarrow &&		\int_{-\infty}^{\infty}\text{d}t \, e^{i\omega t} D^{>}(t) = \int_{-\infty}^{\infty}\text{d}t \, e^{i\omega t} D^{<}(-t) = \int_{-\infty}^{\infty}\text{d}t \, e^{i\omega t} D^{<}(t+i/T) \nonumber\\
    \Rightarrow &&	D^{>}(\omega) = D^{<}(-\omega) = e^{\omega/T} D^{<}(\omega) \nonumber\\
    \Rightarrow &&		\left. \frac{\text{d} D^{>}(\omega)}{\text{d}\omega} \right|_{\omega=0} = \left. -\frac{\text{d} D^{<}(\omega)}{\text{d}\omega} \right|_{\omega=0} = \frac{1}{T} D^{<}(\omega=0) + \left. \frac{\text{d} D^{<}(\omega)}{\text{d}\omega} \right|_{\omega=0} \nonumber\\
    \Rightarrow &&		\left. \frac{\text{d} D^{>}(\omega)}{\text{d}\omega} \right|_{\omega=0} = \left. -\frac{\text{d} D^{<}(\omega)}{\text{d}\omega} \right|_{\omega=0} = \frac{1}{2T} D^{<}(\omega=0).
\end{eqnarray}
Inserting the last equality into eq.~(\ref{eq:postulate}), we have
\begin{equation}\label{eq:intermediatestep}
	i
	\int_{0}^{\infty}\text{d}t \, t \Big \langle \tilde{E}^{a}_i(t,\vec{0})\tilde{E}^{a}_i(0,\vec{0}) \Big \rangle =
	\frac{1}{4T} D^{<}(\omega=0). 
\end{equation}
Taking $\kappa$ as written in eq.~(\ref{eq:kappadefinition}), we can bring it to the form
\begin{eqnarray}
		\kappa &=& \frac{g^{2}}{6 N_{c}}\int^{+\infty}_{0} \text{d}t \, \Big \langle \left\{ \tilde{E}^{a}_i(t,\vec{0}),\tilde{E}^{a}_i(0,\vec{0}) \right\} \Big \rangle 
		= \frac{g^{2}}{6 N_{c}} 
		\int^{+\infty}_{0} \text{d}t \, \left( D^{>}(t) + D^{<}(t) \right) \nonumber\\
		&=& \frac{g^{2}}{6 N_{c}} \int^{0}_{-\infty} \text{d}t \, D^{<}(t) + \frac{g^{2}}{6 N_{c}} \int^{+\infty}_{0} \text{d}t \, D^{<}(t) 
		= \frac{g^{2}}{6 N_{c}} \int^{\infty}_{-\infty} \text{d}t \, D^{<}(t) \nonumber\\
		&=& \frac{g^{2}}{6 N_{c}}  D^{<}(\omega=0).
\end{eqnarray}
Finally, from eq.~(\ref{eq:intermediatestep}), we obtain  an expression for the integral of the chromoelectric correlator times one power of $t$ in terms of the transport coefficient $\kappa$
\begin{equation}
	i\frac{g^{2}}{6 N_{c}}\int_{0}^{\infty}\text{d}t \, t \Big \langle \tilde{E}^{a}_i(t,\vec{0})\tilde{E}^{a}_i(0,\vec{0}) \Big \rangle = \frac{\kappa}{4T}.
\end{equation}

\section{Conditions for obtaining a Lindblad equation}\label{app:nlo_lindblad}
We aim to write a Lindblad equation equivalent to the master equation given in eq.~\eqref{eq:master_equation}.
There are three distinct vector spaces labeling color, spatial direction, and the transition operator.
The color vector space labels the entries of the transition operator matrices, i.e., $(L_{i}^{n})_{ab}$ where $a$ and $b$ take the values $0$ and $1$.
The spatial direction is labeled by the index $i$ taking the values $1$, $2$, and $3$.
The transition operators are also labeled by the indices $m$ or $n$ taking the values from $0$ to $3$.

Because the matrix $h$ is block diagonal (see eq.~\eqref{eq:metric_tensor}), the problem of reducing the master equation given in eq.~\eqref{eq:master_equation} into Lindblad form becomes the problem of finding a collapse operator for each of the blocks.
Let us  consider, for instance, the block elements $h_{00}=h_{11}=0$ and $h_{10}=h_{01}=1$.
The collapse operator $C$ we seek must satisfy 
\begin{equation}\label{eq:octet_master_equation}
	\sum_{n,m=0}^{1} h_{nm} \left( L^{n} \rho(t) L^{m\,\dagger} - \frac{1}{2} \left\{ L^{m\,\dagger} L^{n}, \rho(t)  \right\} \right)
=  C \rho(t) C^{\dagger} - \frac{1}{2} \left\{ C^{\dagger} C, \rho(t) \right\}.
\end{equation}
It therefore exists if the equation
\begin{equation}\label{eq:for_collapse_operators}
	\begin{pmatrix} L^{0\,\dagger} & L^{1\,\dagger} \end{pmatrix} 
	\begin{pmatrix} 0 & 1 \\ 1 & 0 \end{pmatrix} 
	\begin{pmatrix} L^{0} \\ L^{1} \end{pmatrix} 
	= L^{0\,\dagger} L^{1} + L^{1\,\dagger} L^{0} 
	= C^{\dagger} C 
\end{equation}
has a solution. 
An equivalent condition defines a second collapse operator 
from $L^2$ and $L^3$, when considering the block elements $h_{22}=h_{33}=0$ and $h_{23}=h_{32}=1$.

We investigate under which circumstances eq.~\eqref{eq:for_collapse_operators} has a solution.
First, we examine the case when $L^{0}$ and $L^{1}$ are linearly dependent, i.e., 
\begin{equation}
\label{eq:Llindep}
L^{1} = (a+ib)L^{0}, 
\end{equation}
with $a$ and $b$ real numbers. 
This is the case at LO in the $E/T$ expansion, where both $L^0$ and $L^1$ are proportional to $r_i$.
Inserting eq.~\eqref{eq:Llindep} into eq.~\eqref{eq:for_collapse_operators}, we obtain 
\begin{equation}
2a L^{0\,\dagger}L^{0} = C^{\dagger}C,
\end{equation}
the solution of which is 
\begin{equation}\label{eq:leading_order_collapse_operator}
C = \sqrt{2a} L^{0}.
\end{equation}
When rewriting this result in matrix form starting from the left-hand side of eq.~\eqref{eq:for_collapse_operators},
\begin{equation}
    \begin{pmatrix} L^{0\,\dagger} & (a-ib)L^{0\,\dagger} \end{pmatrix}
	\begin{pmatrix} 0 & 1 \\ 1 & 0 \end{pmatrix}
	\begin{pmatrix} L^{0} \\ (a+ib)L^{0} \end{pmatrix} = 
	\begin{pmatrix} 0 & \sqrt{2a}L^{0\,\dagger} \end{pmatrix}
	\begin{pmatrix} -1 & 0 \\ 0 & 1 \end{pmatrix}
	\begin{pmatrix} 0 \\ \sqrt{2a} L^{0} \end{pmatrix},\label{eq:diagonalization_2}
\end{equation}
one may rotate the vector $(L^0,L^1)$ in such a way that it becomes orthogonal to the eigenspace of the matrix $h_{nm}$ with negative eigenvalue.

We now examine the case when $L^{0}$ and $L^{1}$ are neither linearly dependent 
nor orthogonal, i.e., 
\begin{equation}\label{eq:nlo_l1}
	L^{1} = (a+ib)L^{0} + \epsilon L^{1}_{(1)},
\end{equation}
where $\epsilon$ is a complex scalar and $L^{1}_{(1)}$ is nonzero and linearly independent of $L^{0}$, i.e., there exists no scalar $c$ for which $L^{0} = c L^{1}_{(1)}$.
This is the case at NLO in the $\epsilon = E/T$ expansion, where $L^1$ contains both a term proportional 
to $r_i$, and hence to $L_0$, and a term proportional to $p_i$ (see eq.~\eqref{eq:Li1NLO}).
In this case, the left-hand side of eq.~\eqref{eq:for_collapse_operators} can be written as 
\begin{align}
    	&\begin{pmatrix} L^{0\,\dagger} & \left[(a-ib)L^{0\,\dagger} + \epsilon^{*} L^{1\,\dagger}_{(1)}\right] \end{pmatrix}
	\begin{pmatrix} 0 & 1 \\ 1 & 0 \end{pmatrix}
	\begin{pmatrix} L^{0} \\ \left[(a+ib)L^{0} + \epsilon^{*} L^{1}_{(1)}\right] \end{pmatrix} = 
    \nonumber\\
    & \hspace{2cm}
   	\begin{pmatrix} \frac{\epsilon^{*}}{\sqrt{2a}} L^{1\,\dagger}_{(1)} & \left[\sqrt{2a} L^{0\,\dagger} + \frac{\epsilon^{*}}{\sqrt{2a}} L^{1\,\dagger}_{(1)}\right] \end{pmatrix}
	\begin{pmatrix} -1 & 0 \\ 0 & 1 \end{pmatrix}
	\begin{pmatrix} \frac{\epsilon}{\sqrt{2a}} L^{1}_{(1)} \\ \left[\sqrt{2a} L^{0} + \frac{\epsilon}{\sqrt{2a}} L^{1}_{(1)}\right] \end{pmatrix}. 
\end{align}
This is not of Lindblad form as the negative eigenvalue of the matrix $h_{nm}$ contributes through the vector component $\epsilon L^{1}_{(1)}/\sqrt{2a}$. However, in the context of 
an expansion in $\epsilon$, 
discounting this component affects the master equation only at order~$\epsilon^2$.
Hence, if we aim to obtain a master equation that is accurate to order $\epsilon$, it holds that 
\begin{equation}
    	\begin{pmatrix} L^{0\,\dagger} & \left[(a-ib)L^{0\,\dagger} + \epsilon^{*} L^{1\,\dagger}_{(1)}\right] \end{pmatrix}
	\begin{pmatrix} 0 & 1 \\ 1 & 0 \end{pmatrix}
	\begin{pmatrix} L^{0} \\ \left[(a+ib)L^{0} + \epsilon^{*} L^{1}_{(1)}\right] \end{pmatrix} 
  = C^\dagger C + O(\epsilon^2), 
\end{equation}
with collapse operator
\begin{equation}\label{eq:nlo_collapse_operator}
	C = \sqrt{2a} L^{0} + \frac{\epsilon}{\sqrt{2a}} L^{1}_{(1)}.
\end{equation}
Equivalently, one can solve eq.~\eqref{eq:for_collapse_operators} for $L^{1}$ as given in eq.~\eqref{eq:nlo_l1}, i.e.,
\begin{align}
	L^{0\,\dagger} \left[(a+ib) L^{0} + \epsilon L^{1}_{(1)}\right] + \left[(a-ib) L^{0\,\dagger} + \epsilon^{*} L^{1\,\dagger}_{(1)}\right] L^{0} 
	&= 2a L^{0\,\dagger}L^{0} + \epsilon L^{0\,\dagger}L^{1}_{(1)} 
	+ \epsilon^{*} L^{1\,\dagger}_{(1)}L^{0} \nonumber\\
	&= C^\dagger C,
\end{align}
whose solution is eq.~\eqref{eq:nlo_collapse_operator} up to corrections of order 
$\epsilon^2$ to $C^\dagger C$.

\section{Lindblad equation in the spherical basis}
\label{app:spherical}

In this section, we examine the angular momentum structure of the Lindblad equation.
We define the density matrix projected onto states of definite azimuthal quantum number $l$ and magnetic quantum number $m$ as
\begin{equation}
	\rho^{l'm';lm} = \langle l'm' | \rho | lm \rangle = \int \text{d}\Omega\, Y_{l'm'*}(\theta, \phi)\, \rho\, Y_{lm}(\theta, \phi) \, .
\end{equation}
Since the density matrix is radially symmetric, i.e., it possesses no preferred direction, it is diagonal in $l$ and $m$, i.e.,
\begin{equation}
	\rho^{l'm';lm} = \rho^{lm;lm}\delta_{ll'}\delta_{mm'} \, .
\end{equation}
Radial symmetry, furthermore, implies equal probability for all polarizations within an orbital.  
All information can, therefore, be encoded in
\begin{equation}\label{eq:rho_l_definition}
	\rho^{l} = \sum_{m} \rho^{lm;lm},\quad \rho^{lm;lm}=\frac{1}{2l+1}\rho^{l} \, .
\end{equation}
We note that for our calculation (both at LO and NLO), the Lindblad equation contains only vector operators. Given a Lindblad equation with a vector jump operator ($\Vec{C} = \{C_{x}, C_{y}, C_{z}\}$), one can first go to the spherical basis given by the operators $\{C^{(1)}_{+1},C^{(1)}_{0},C^{(1)}_{-1}\}$, defined as 
\begin{equation}
	C^{(1)}_{\pm 1} = \mp \frac{1}{\sqrt{2}} \left( C_{x} \pm i C_{y} \right) , \quad
	C^{(1)}_{0} = C_{z} \, ,
\end{equation}
where, anticipating the use of the Wigner--Eckart theorem, we explicitly label the rank of the spherical tensor operator with the superscript $(1)$.
An inner product of the form $C^{\dagger}_{i}C_{i}$ is written the same in the Cartesian and spherical bases, i.e.,
\begin{equation}
	\sum_{i} C_{i}^{\dagger} C_{i} = \sum_{q} C^{(1)\dagger}_{q}C^{(1)}_{q}\quad,\quad \sum_{i}C_{i}\rho \, C^{\dagger}_{i} = \sum_{q} C^{(1)}_{q}\rho \, C^{(1)\dagger}_{q} \, ,
\end{equation}
where it should be noted that $C^{(1)}_{+1} = -(C^{(1)}_{-1})^{\dagger}$ and  $C^{(1)}_{-1} = -(C^{(1)}_{+1})^{\dagger}$.

We start by projecting the Lindblad equation onto the spherical basis. The terms containing scalar operators (such as $\text{Re}[H^{\mathrm{eff}}_{s/o}]$ and $\text{Im}[H^{\mathrm{eff}}_{s/o}]$) are easy to project
\begin{align}
    \sum_{m} \bra{lm} S \rho \ket{lm} &= \sum_{m,m',l'} \bra{lm} S \ket{l'm'} \bra{l'm'} \rho \ket{lm} \nonumber \\
     &= \sum_{m} \bra{lm} S \ket{lm} \rho^{lm;lm} \nonumber \\
     &=  S\rho_{l} \, ,
\end{align}
where in the second line we have used $\rho^{lm;l'm'} = \rho^{lm;lm}\delta_{ll'}\delta_{mm'}$ and in the last line we have used $\bra{lm} S \ket{lm} = S$, which is true for any scalar operator $S$ that is a function of radial operators only. As a result, we obtain for the scalar terms
\begin{align}
    \label{eq: from cartesian to spherical Lindblad}
    \sum_{m} \bra{lm}\frac{\mathrm{d}\rho(\Vec{r},\Vec{r}^{\,\prime},t)}{\mathrm{d}t}\ket{lm} & = \frac{\mathrm{d}\rho_{l}(r,r',t)}{\mathrm{d}t} \nonumber, \\
    \sum_{m} \bra{lm} H^{\mathrm{eff}}_{s/o},\rho(\Vec{r},\Vec{r}',t) \ket{lm} &= H^{\mathrm{eff}}_{s/o} \rho_{l}(r,r',t) \, .
\end{align}

The only non-trivial calculation is that of the term $C_{i}\rho C^{\dagger}_{i}$ as the operators $C_{i}$ are vector operators and give rise to transitions between different states in color and angular momentum. 
The evaluation of this term can be done using the Wigner--Eckart theorem. 
The Wigner--Eckart theorem states that we can write the angular momentum matrix element of a vector operator $C^{(1)}_{q}$ as
\begin{equation}\label{eq:wigner_eckart}
	\langle l^\prime m^\prime |C^{(1)}_{q}|lm\rangle = \langle lm;1q|l^\prime m^\prime \rangle \langle l'||C^{(1)}||l\rangle \, .
\end{equation}
The advantage of using the spherical basis is clear now. The angular matrix element of the operators $C^{(1)}_{0,\pm 1}$ can be expressed as Clebsch--Gordon coefficients $\langle lm;1q|l^\prime m^\prime \rangle$
multiplied by the reduced matrix element $\langle l'|| C^{(1)} ||l\rangle$.  

\subsection{Projecting the jump term \texorpdfstring{$C_{i}\rho C^{\dagger}_{i}$}{Ci rho Cdagi} on the spherical basis}

For the calculation of the jump term, we begin by projecting onto the eigenstate $\ket{lm}$
\begin{align}\label{eq:projection}
	\sum_{m,q} \bra{lm} C^{(1)}_{q}\rho C^{(1)\,\dagger}_{q} \ket{lm} 
	&= \sum_{\substack{q,m,l_{1},m_{1},\\l_{2},m_{2}}}  \bra{lm} C^{(1)}_{q} \ket{l_{1}m_{1}}\bra{l_{1}m_{1}} \rho \ket{l_{2}m_{2}} \bra{l_{2}m_{2}} C^{(1)\,\dagger}_{q} \ket{lm} \nonumber \\
	& \hspace{-2cm}= \sum_{q,m,l_{1},m_{1}}  \bra{lm} C^{(1)}_{q} \ket{l_{1}m_{1}} \rho^{l_{1}m_{1};l_{1}m_{1}}  \bra{l_{1}m_{1}} C^{(1)\,\dagger}_{q} \ket{lm} \nonumber \\
	&\hspace{-2cm}= \sum_{q,m,l_{1},m_{1}} \bra{l_{1}m_{1}; 1q}lm\rangle \langle l|| C^{(1)}|| l_{1} \rangle \rho^{l_{1}m_{1};l_{1}m_{1}} \braket{lm;1q|l_{1}m_{1}} \langle l_{1}|| C^{(1)\,\dagger} || l \rangle.
\end{align}
The Lindblad operators at NLO can be parametrized as $	C_{i} = a r_{i} + b p_{i}$ where $a$ and $b$ are radially symmetric complex matrix operators. 
To proceed further, we need to evaluate the 
reduced matrix elements $\langle l'|| C^{(1)} || l \rangle$ for the two cases separately, $C_{i} = r_{i}, p_{i}$. 
We present the details for each of these calculations in the next subsections.

\subsection{Calculation of \texorpdfstring{$\langle l' || r^{(1)} || l\rangle$}{<l'||r||l>}}

To calculate $\langle l' || r^{(1)} || l\rangle$, we make use of the Wigner--Eckart relation. 
To evaluate the reduced matrix element $\langle l' || r^{(1)} || l \rangle$, we consider the case of $q=m=m^\prime=0$, i.e., from eq.~(\ref{eq:wigner_eckart}), we have 
\begin{equation}\label{eq:alternate_wigner_eckart}
	\langle l' || r^{(1)} || l \rangle  = \frac{\langle l' 0 | z | l 0 \rangle}{\langle l0; 10 | l' 0 \rangle },
\end{equation}
where $z = r \cos{\theta}$ can be expressed in terms of spherical harmonics using $\cos\theta = \sqrt{(4\pi)/3} \, Y_{10}$. By using the following identities for the spherical harmonic functions and Clebsch--Gordon coefficients,
\begin{equation}
	\int \text{d} \, \Omega \, Y_{l_{1}m_{1}} \, Y_{l_{2}m_{2}} \, Y^{*}_{l_{3}m_{3}} 
	= \sqrt{\frac{(2l_{1}+1)(2l_{2}+1)}{4\pi(2l_{3}+1)}} \langle l_{1}m_{1} ;\, l_{2}m_{2} | l_{3}m_{3} \rangle \langle l_{1}0 ;\, l_{2}0 | l_{3}0 \rangle \, ,
\end{equation}
and
\begin{equation} \label{eq:l010lp0}
	\langle l0 ;\, 10 | l'0 \rangle = \sqrt{\frac{l+1}{2l+1}} \delta_{l',l+1} - \sqrt{\frac{l}{2l+1}} \delta_{l',l-1} \, ,
\end{equation}
we obtain
\begin{equation}
    \label{eq:reduced r optr}
    \langle l' || r^{(1)} || l\rangle =
	\frac{\langle l' 0 | r \cos\theta | l  0 \rangle}{\langle l0; 10| l' 0 \rangle}  = 
	\sqrt{\frac{l+1}{2l+3}} \delta_{l',l+1} r - \sqrt{\frac{l}{2l-1}} \delta_{l',l-1} r \, .
\end{equation}

\subsection{Calculation of \texorpdfstring{$\langle l' || p^{(1)} || l\rangle$}{<l'||p||l>}}

To evaluate the reduced momentum operator $\langle l' || p^{(1)} || l \rangle$ we need to evaluate
\begin{equation}\label{eq:alternate_wigner_eckart2}
	\langle l' || p^{(1)} || l \rangle  = \frac{\langle l' 0 | p_z | l 0 \rangle}{\langle l0; 10 | l' 0 \rangle } \, .
\end{equation}
The momentum operator $p_{z} = -i\partial_z$ in spherical coordinates is given by 
\begin{equation}\label{eq:p_0 in radial coord}
	p_{z} = -i\left[ \cos \theta \, \partial_r - \frac{\sin \theta}{r} \partial_\theta \right].
\end{equation}

The calculation of the term $\cos{\theta}\,\partial_{r}$ follows that of $ r \cos\theta$ and reduces to
\begin{equation}\label{eq:cos_theta}
	\frac{\langle l' 0 | \cos\theta \partial_{r} | l  0 \rangle}{\langle l0; 10| l' 0 \rangle}  = 
	\sqrt{\frac{l+1}{2l+3}} \delta_{l',l+1}\,\partial_{r}- \sqrt{\frac{l}{2l-1}} \delta_{l',l-1}\,\partial_{r} \, .
\end{equation}
Evaluating the second term using the relation 
\begin{equation}
	Y^{*}_{l0}(\theta, \phi) = Y_{l0}(\theta, \phi) = \sqrt{\frac{2l+1}{4\pi}} P_{l}(\cos\theta),
\end{equation}
one finds
\begin{align}
	\langle l' 0 | \sin\theta \partial_\theta | l0 \rangle
	&= \int \text{d} \Omega \, Y_{l0}^{*}(\theta, \phi) \sin\theta \, \partial_\theta \,  Y_{l0}(\theta, \phi) \nonumber \\
	&= \frac{1}{2} \sqrt{(2l'+1)(2l+1)} \int_{-1}^{-1} \text{d}x\, P_{l'}(x)(x^{2}-1)P'_{l}(x) \, ,
\end{align}
where $x=\cos\theta$ and the prime indicates a derivative with respect to the argument of the Legendre polynomial.
We make use of a number of orthogonality and recurrence properties of the Legendre polynomials
\begin{align}
	(x^{2}-1)P'_{l} &= l \left[ x P_{l}(x) - P_{l-1}(x) \right]\nonumber \\
	&= l \left[ P_{1}(x) P_{l}(x) - P_{l-1}(x) \right], \\
	\int_{-1}^{1}\text{d}x\, P_{l_{1}}(x)P_{l_{2}}(x)P_{l_{3}}(x) &= \frac{2}{2l_{3}+1} \langle l_{1}0;\, l_{2}0 | l_{3}0 \rangle^{2},\\
	\int_{-1}^{1}\text{d}x\, P_{l_{1}}(x)P_{l_{2}}(x) &= \frac{2}{2l_{1}+1} \delta_{l_{1}l_{2}} \, ,
\end{align}
to write
\begin{equation}
	\frac{1}{2} \sqrt{(2l'+1)(2l+1)} \int_{-1}^{-1} \text{d}x\, P_{l'}(x)(x^{2}-1)P'_{l}(x)
	= l\sqrt{\frac{2l+1}{2l'+1}} \left( \langle l0;\, 10 | l'0 \rangle^{2} - \delta_{l',l-1} \right) .
\end{equation}
Using eq.~(\ref{eq:l010lp0}), this becomes
\begin{equation}
	\langle l' 0 | \sin\theta \, \partial_\theta | l  0 \rangle
	= l\frac{l+1}{\sqrt{(2l+3)(2l+1)}} \delta_{l',l+1} - l\frac{l+1}{\sqrt{(2l-1)(2l+1)}} \delta_{l',l-1} \, ,
\end{equation}
and, furthermore,
\begin{equation}\label{eq:sin_theta_d_theta}
	\frac{\langle l' 0 | \sin\theta \,\partial_\theta | l  0 \rangle}{\langle l0; 10| l' 0 \rangle}
	= l\sqrt{\frac{l+1}{2l+3}}\delta_{l',l+1} + (l+1) \sqrt{\frac{l}{2l-1}}\delta_{l',l-1} \, .
\end{equation}
Combining eqs.~(\ref{eq:alternate_wigner_eckart2}), (\ref{eq:p_0 in radial coord}), (\ref{eq:cos_theta}), and (\ref{eq:sin_theta_d_theta}), we obtain
\begin{align}
\label{eq: reduced momentum optr}
	\langle l' || p^{(1)} || l \rangle =& \frac{\displaystyle \langle l^\prime 0 |\left[-i \left( \cos\theta \,\partial_r-\frac{\sin\theta}{r}\,\partial_\theta\right)\right]|l0\rangle}{ \langle l0;10|l^\prime 0 \rangle} \nonumber \\
	=& \hphantom{{}-{}}\sqrt{\frac{l+1}{2l+3}} \left[-i \left( \partial_r - \frac{l}{r} \right) \right] \delta_{l',l+1} 
	-\sqrt{\frac{l}{2l-1}} \left[-i \left( \partial_r + \frac{l+1}{r} \right) \right] \delta_{l',l-1} \, 
\end{align}

We note that the reduced matrix elements $\langle l \pm 1|| p^{(1)} || l \rangle$ are not Hermitian.
To derive the Hermitian conjugate of the $p^{\uparrow /\downarrow}$ reduced matrix elements, one can use the following result for the Hermitian conjugate of the partial derivative operator $-i\partial_{r}$ in the radial basis
\begin{equation}
    \label{eq: hermitian conjugate of partial_r}
    \langle\psi|-i\partial_{r}|\phi\rangle = \int \mathrm{d}r \ r^{2} \psi^{*}(r)[-i\partial_{r}\phi(r)] = \int \mathrm{d}r \ r^{2} \left[ -i\left( \partial_{r} + \frac{2}{r}\right)\psi(r)\right]^{*}\phi(r) \, ,
\end{equation}
which gives $(-i\partial_{r})^{\dagger} = -i\left(\partial_{r}+ 2/r \right)$. 

We use the following identity for simplifying sums over $q$ and $m$  (which can be shown for any general vector operator $C^{(1)}$)
\begin{align}\label{eq: sum over clb coeff jump term}
    \sum_{m, q} \bra{lm}C^{(1)}_{q}\rho(\Vec{r},\Vec{r}^{\,\prime},t)(C^{(1)}_{q})^{\dagger}&\ket{lm} =  \frac{2l-1}{2l+1} C^{\downarrow} \rho^{l-1}(r,r',t)C^{\downarrow\,\dagger} + \frac{2l+3}{2l+1} C^{\uparrow} \rho^{l+1}(r,r',t)C^{\uparrow\,\dagger}
\end{align}
where $C^{\uparrow,\downarrow} = \langle l \pm 1 || C^{(1)} || l \rangle$.
Finally, using eqs.~\eqref{eq:reduced r optr}, \eqref{eq: reduced momentum optr} and \eqref{eq: sum over clb coeff jump term}, we obtain
\begin{align}
    & \sum_{m, i} \bra{lm}(a r_{i} + b p_{i})\rho(\Vec{r},\Vec{r}^{\,\prime},t)(a r_{i} + b p_{i})^{\dagger}\ket{lm} = \nonumber \\
    & \hspace{3cm} \frac{l+1}{2l+1}\left[ a r - b i \left( \partial_r - \frac{l}{r} \right) \right] \rho^{l-1} 
	\left[ a^{\dagger} r - b^{\dagger} i \left( \partial_r + \frac{l+2}{r} \right) \right] \nonumber\\
	&\hspace{3cm} + \frac{l}{2l+1}\left[ a r - b i \left( \partial_r + \frac{l+1}{r} \right) \right] \rho^{l+1} 
	\left[ a^{\dagger} r - b^{\dagger} i \left( \partial_r - \frac{l-1}{r} \right) \right] .
\end{align}

In summary, we have shown that starting from the three-dimensional Lindblad equation given in eq.~\eqref{eq:Lindblad} and using the rotational symmetry present in the problem one can reduce it to a one-dimensional Lindblad equation with operators acting only on the radial part of the wave function.

\subsection{Projection into the reduced radial space}

A further simplification of both the Lindblad operators and required radial differential operators is possible when one projects them into the reduced wave function basis ($u$-space). 
We have already noted that in eq.~\eqref{eq: hermitian conjugate of partial_r} the derivative operator is not Hermitian.
In numerical implementations this creates a problem since it is not straightforward to implement such a non-Hermitian derivative on a one-dimensional lattice. 
To fix this problem, as is familiar in the construction of a one-dimensional effective Hamiltonian for the hydrogen atom, we redefine the radial wave function $\psi(r) = u(r)/r$, where $u(r)$ is the reduced wave function.  We then project all operators into the reduced wave function basis ($u$-space), i.e.,
\begin{equation}
    \int \mathrm{d}r \ r^{2} \psi^{*}(r) \hat{A} \phi(r) \equiv \int \mathrm{d}r \
    u_{\psi}^*(r) \hat{A}_{u} u_{\phi}(r) \, .
\end{equation}

For example, the radial derivative operator $-i\partial_{r}$, which is not Hermitian in the radial wave function basis, is Hermitian in the reduced wave function basis ($u$-space)
\begin{eqnarray}
&& \int \mathrm{d}r \ r^{2} \psi^{*}(r) \left[ -i\partial_{r}\phi(r) \right] = \int \mathrm{d}r \ r^{2} \left[-i\left(\partial_{r}+\frac{2}{r}\right)\psi(r)\right]^{*}\phi(r) \, , \nonumber \\
\Rightarrow && \int \mathrm{d}r \ r^{2} \frac{u_{\psi}^{*}(r)}{r} \left[-i\partial_{r}\frac{u_{\phi}(r)}{r} \right] = \int \mathrm{d}r \ r^{2} \left[-i\left(\partial_{r}+\frac{2}{r}\right)\frac{u_{\psi}(r)}{r} \right]^{*} \frac{u_{\phi}(r)}{r} \, , \nonumber \\
\Rightarrow && \int \mathrm{d}r \ u_{\psi}^{*}(r)\left[-i\left(\partial_{r}-\frac{1}{r}\right)u_{\phi}(r)\right]  = \int \mathrm{d}r \ \left[-i\left(\partial_{r}+\frac{1}{r}\right)u_{\psi}(r)\right]^{*} u_{\phi}(r) \, ,
\end{eqnarray}
which gives $[-i(\partial_{r}+{1}/{r})]^{\dagger} = -i(\partial_{r}-{1}/{r}) \Rightarrow (-i\partial_{r})^\dagger = -i\partial_{r}$ in $u$-space. We denote the spatial derivative in $u$-space as $p_{r}$ to not cause any confusion. The following operators have different forms in the reduced wave function basis ($u$-space) compared to their counterparts in the radial basis ($\psi$-space)
\begin{align}
    \frac{\Vec{p}^{\,2}}{M} &\rightarrow   - \frac{1}{M} \partial_r ^2 + \frac{l(l+1)}{M r^2} = \frac{\mathcal{\overline{D}}^2}{M} \, , \\
    \sum_{i}\{p_{i} , r_{i} \} &\rightarrow  p_{r}r+r p_{r} \, ,  \\
    -i\sqrt{\frac{l+1}{2l+1}}\left(\partial_{r} - \frac{l}{r}\right) &\rightarrow -i\sqrt{\frac{l+1}{2l+1}}\left(\partial_{r} - \frac{l+1}{r}\right)  ,  \\
    -i\sqrt{\frac{l}{2l+1}}\left(\partial_{r} + \frac{l+1}{r}\right) &\rightarrow -i\sqrt{\frac{l}{2l+1}}\left(\partial_{r} + \frac{l}{r}\right)  .
\end{align}

\section{Transport coefficient \texorpdfstring{$\kappa$}{kappa}}\label{app:kappa}

As has been pointed out in refs.~\cite{Eller:2019spw,Binder:2021otw,Scheihing-Hitschfeld:2022xqx}, the transport coefficient $\kappa$ that enters into the quarkonium evolution equation and the heavy quark momentum diffusion coefficient computed on the lattice differ in the time ordering of the fields in the correlator. 
They both vanish in vacuum at all orders in perturbation theory and dimensional regularization, but 
they could differ due to finite temperature effects. Investigations are under way, but no quantitative conclusion has been reached yet, neither in weakly coupled thermal field theory, nor in lattice thermal QCD.
In lattice thermal QCD, the coefficient $\kappa$, as defined in eq.~\eqref{eq:kappadefinition}, has not been calculated directly. 
Indirect determinations based on the pNRQCD definition of the quarkonium thermal width, which is proportional to $\kappa$, and unquenched lattice measurements of the quarkonium thermal width provide a range that fully overlaps with the  direct lattice determinations of the heavy quark momentum diffusion coefficient~\cite{Brambilla:2019tpt}.
In the body of the paper, in the absence of a direct lattice determination of the correlator defining the quarkonium momentum diffusion coefficient, we follow the approach of refs.~\cite{Brambilla:2020qwo,Brambilla:2021wkt} and parameterize the temperature dependence of $\kappa$ based on the lattice determination of the heavy quark momentum diffusion coefficient done in ref.~\cite{Brambilla:2020siz}. 
\begin{figure}[h!]
\begin{center}
\includegraphics[width=0.48\linewidth]{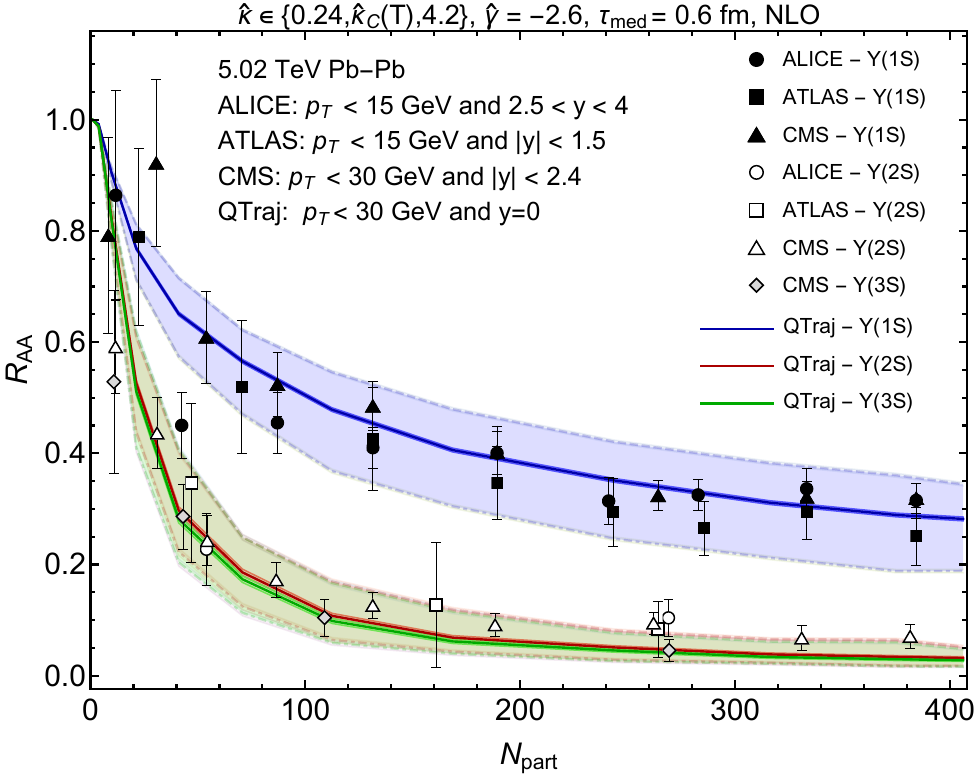} \hspace{3mm}
\includegraphics[width=0.48\linewidth]{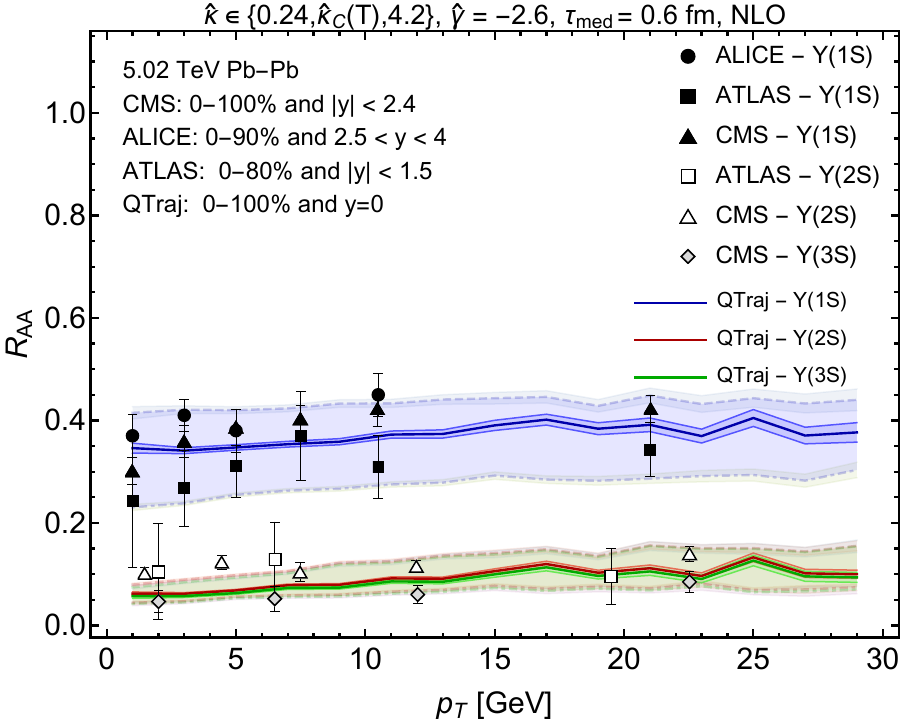}
\end{center}
\caption{
    Predictions for $\Upsilon(1S)$, $\Upsilon(2S)$, and $\Upsilon(3S)$ $R_{AA}$ as a function of $N_{\rm part}$ (left) and $p_{T}$ (right) with $\tau_\text{med} = 0.6$ fm, NLO evolution, and $T_{f} = 190$ MeV.  
    In both panels, the lower, central, and upper curves correspond to $\hat\kappa = \{ 0.24, \kappa_C(T), 4.2 \}$ with $\hat\gamma=-2.6$.
}
\label{fig:constant_kappa}
\end{figure}

Using only indirect lattice determinations of $\kappa$ would not significantly change the central value of our results but somewhat enlarges the uncertainties due to the variation of $\kappa$.
In fig.~\ref{fig:constant_kappa}, we estimate these errors by running simulations with $\hat{\kappa} = 0.24$ and $4.2$ (with $\hat{\gamma}=2.6$) as taken from eq.~(36) of ref.~\cite{Brambilla:2019tpt}.
We observe larger uncertainty compared to the $\hat{\kappa}$ variation presented in figs.~\ref{fig:raavsnpart1} and \ref{fig:raavspt1}, but, even with this larger $\hat{\kappa}$ variation, we find smaller uncertainty compared to the $\hat{\gamma}$ variation.

\section{Dependence on the medium initialization time}
\label{app:taumed}

\begin{figure}[h!]
\begin{center}
\includegraphics[width=0.48\linewidth]{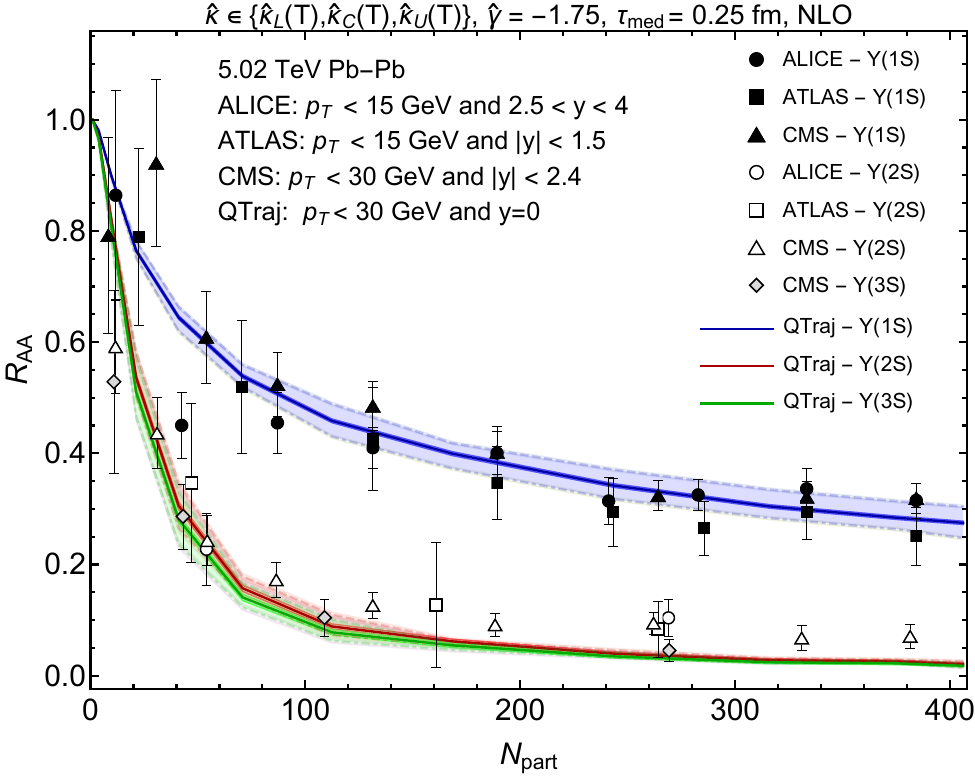} \hspace{3mm}
\includegraphics[width=0.48\linewidth]{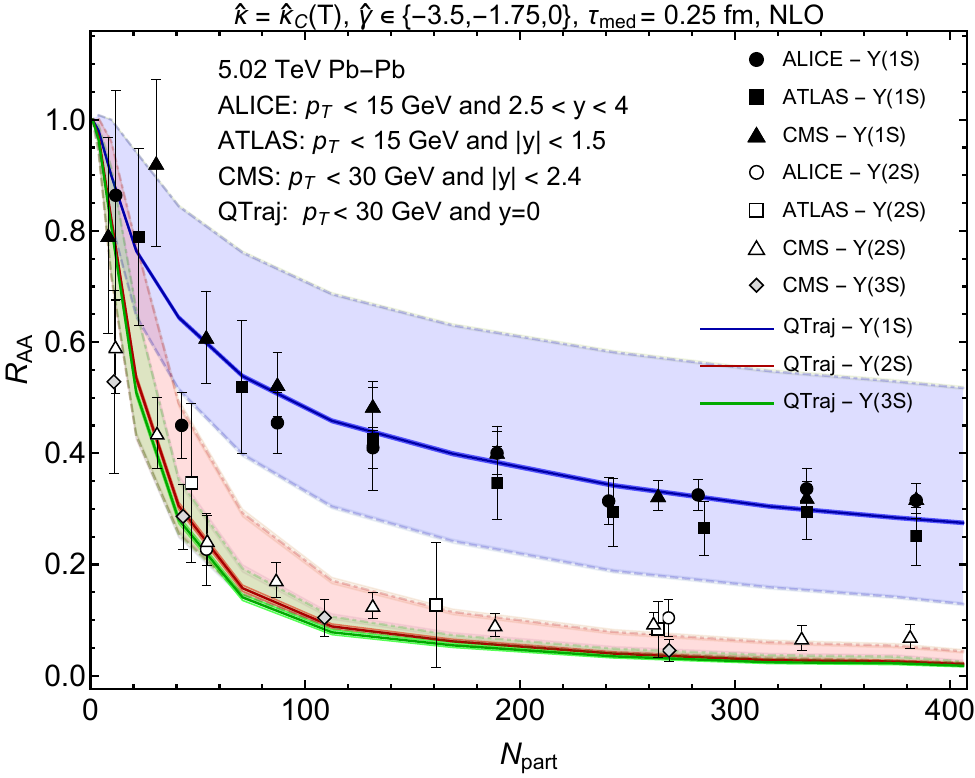}
\end{center}
\caption{$R_{AA}$ predictions for the $\Upsilon(1S)$, $\Upsilon(2S)$, and $\Upsilon(3S)$ as a function of $N_{\rm part}$ with $\tau_\text{med} = 0.25$ fm.  The left panel shows variation of $\hat\kappa \in \{ \kappa_L(T), \kappa_C(T), \kappa_U(T) \}$ and the right panel shows variation of $\hat\gamma$ in the range $-3.5 \leq \hat\gamma \leq 0$.  In both panels, the solid line corresponds to $\hat\kappa = \hat\kappa_C(T)$ and $\hat\gamma = -1.75$.  The experimental measurements shown are from the ALICE~\cite{Acharya:2020kls}, ATLAS~\cite{ATLAS5TeV}, and CMS~\cite{Sirunyan:2018nsz,CMSupsilonQM2022} collaborations.}
\label{fig:raavsnpart2}
\end{figure}

\begin{figure}[h!]
\begin{center}
\includegraphics[width=0.48\linewidth]{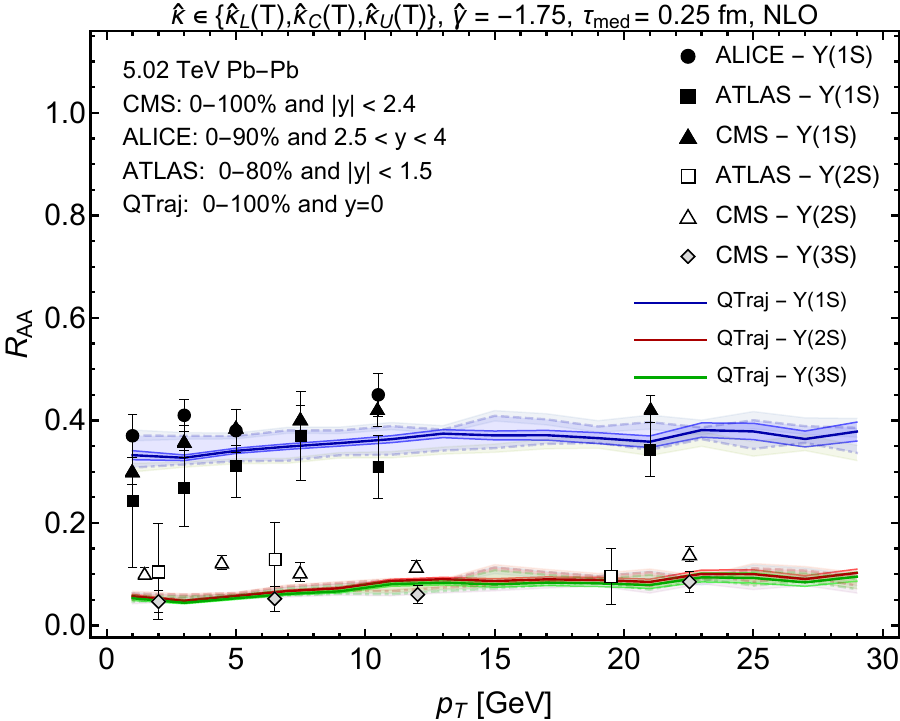} \hspace{3mm}
\includegraphics[width=0.48\linewidth]{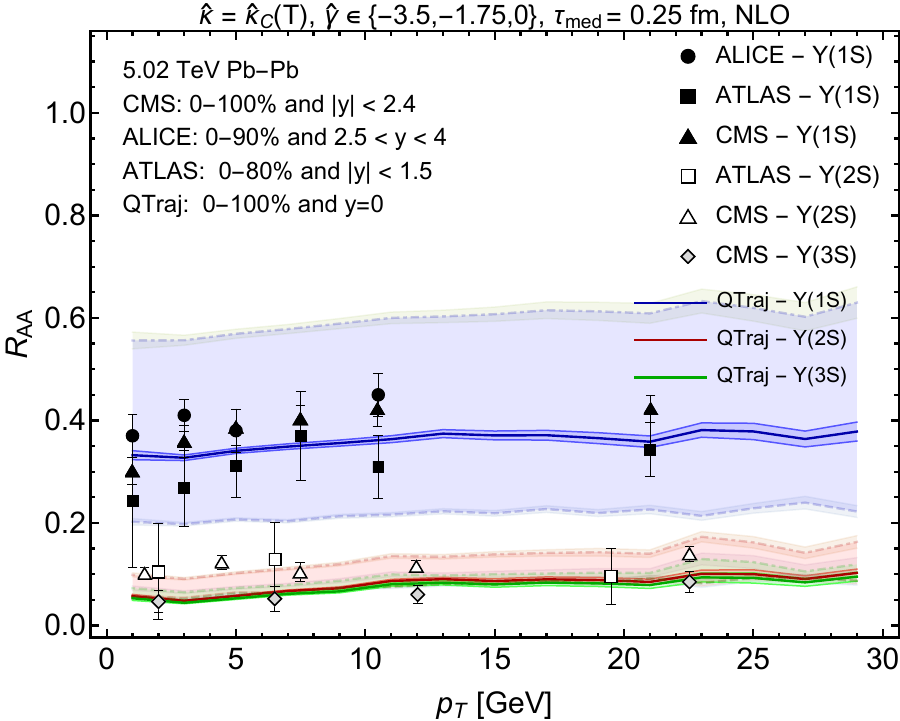}
\end{center}
\caption{
$R_{AA}$ for the $\Upsilon(1S)$, $\Upsilon(2S)$, and $\Upsilon(3S)$ as a function of $p_T$.  The value of $\tau_{\rm med}$, the meaning of the bands, and experimental data sources are the same as in fig.~\ref{fig:raavsnpart2}.
}
\label{fig:raavspt2}
\end{figure}

In this appendix, we consider the dependence of our results on the assumed medium initialization time $\tau_{\rm med}$.  In the body of the paper, we assumed $\tau_{\rm med}$ = 0.6 fm.  Here we consider the effect of an early medium initialization time by choosing instead $\tau_{\rm med}$ = 0.25 fm, which is the earliest proper time available in the hydrodynamic simulations.  
In figs.~\ref{fig:raavsnpart2} and \ref{fig:raavspt2}, we present our NLO predictions for $R_{AA}$ as a function of $N_{\rm part}$ and $p_T$, respectively, including the effect of excited state feed down.  
In these figures, the central solid line corresponds to $\hat\gamma = -1.75$ as opposed to the value of $\hat\gamma = -2.6$ used in figs.~\ref{fig:raavsnpart1} and \ref{fig:raavspt1}.  
Due to the shift in the central value of $\hat\gamma$, both choices for $\tau_{\rm med}$ provide a good description of the data.  
From the comparison with the case of $\tau_{\rm med}$= 0.6 fm presented in the main body of the paper, we see that the uncertainty of our inferred central value of $\hat\gamma$ is $\Delta\hat\gamma \sim 1$. 
We note that starting the Lindblad evolution at 0.25 fm introduces additional uncertainties related to the fact that at early times in the QGP evolution the system is far from isotropic thermal evolution \cite{Almaalol:2020rnu,Strickland:2013uga}.  
In the future, it may be possible to use methods such as those presented in ref.~\cite{Dong:2021gnb} to include the effect of early-time momentum-space anisotropies in a systematic fashion.

\bibliographystyle{JHEP}
\bibliography{qtraj-nlo}

\end{document}